\documentclass[10pt]{iopart}
\usepackage{iopams,times,graphicx,subfigure,rangcite}

\def\proofbox{\hfill{\ensuremath\Box}}
\eqnobysec
\def\eqref#1{(\ref{#1})}

\usepackage{ulem}

\makeatletter
\def\journal#1&#2,{\begingroup \let\journal=\dummyjournal
               \it #1\unskip~\bf\ignorespaces #2\rm,\endgroup}
\def\dummyjournal{\errmessage{Reference foul up: nested \journal macros}}

\let\true@epsilon=\epsilon
\let\epsilon=\varepsilon

\def\@#1{{\mathbf{#1}}}

\makeatother

\newdimen\LENB \newdimen\LENW \newdimen\THI
\newdimen\LENWH \newdimen\LENTOT \newcount\N
\def\vbrknlnele#1#2#3{
  \LENB=#1pt \LENW=#2pt \THI=#3pt
  \LENWH=\LENW \divide\LENWH by 2
  \LENTOT=\LENB \advance\LENTOT by \LENW
  \vbox to \LENTOT{
    \vbox to \LENWH{}
    \nointerlineskip
    \vbox to \LENB{\hbox to \THI{\vrule width \THI height \LENB}}
    \nointerlineskip
    \vbox to \LENWH{}
  }}

\def\vbrknln#1{
  \N=#1
  \vcenter{
    \vbox{
      \loop\ifnum\N>0
        \vbox to 4pt{\vbrknlnele{2}{2}{0.1}}
        \nointerlineskip
        \advance\N by -1
      \repeat
  }}}

\def\hbrknlnele#1#2#3{
  \LENB=#1pt \LENW=#2pt \THI=#3pt
  \LENTOT=\LENB \advance\LENTOT by \LENW
  \vcenter{
    \vbox to \THI{
      \hbox to \LENTOT{
        \hfil
        \vrule width \LENB height \THI
        \hfil}
  }}}

\def\hblele{\hbrknlnele{2}{2.2}{0.1}}

\def\hblfil{\cleaders\hbox{$ \m@th \mkern1mu \hblele \mkern1mu $}\hfill} \makeatother

\begin{document}
\title[Vector 2D Long Wave-Short Wave Resonance
Interaction]
{Two-component Analogue of Two-dimensional Long Wave-Short Wave Resonance
Interaction Equations: A Derivation and Solutions}
\author{Yasuhiro Ohta$^1$, Ken-ichi Maruno$^2
\footnote{e-mail: kmaruno@utpa.edu}$ and Masayuki Oikawa$^3$}
\address{$^1$~Department of Mathematics, 
Kobe University, Rokko, Kobe 657-8501, Japan}
\address{$^2$
~Department of Mathematics, 
The University of Texas-Pan American, 
Edinburg, TX 78541}
\address{$^3$
~Research Institute for Applied Mechanics,
Kyushu University, Kasuga, Fukuoka, 816-8580, Japan}
\date{\today}
\def\submitto#1{\vspace{28pt plus 10pt minus 18pt}
     \noindent{\small\rm To be submitted to : {\it #1}\par}}

\begin{abstract}
The two-component analogue of 
two-dimensional long wave-short wave resonance
interaction equations is derived in a physical setting. 
Wronskian solutions 
of the integrable two-component analogue of 
two-dimensional long wave-short wave resonance
interaction equations are presented. 
\par
\kern\bigskipamount\noindent
\today
\end{abstract}

\kern-1.5\bigskipamount
\submitto{\JPA}


\section{Introduction}

Recently, vector soliton equations (or coupled soliton equations)
such as the vector Nonlinear Schr\"odinger (vNLS) equation have received so
much attention in mathematical physics and nonlinear physics. 
Especially, the vNLS equation has been studied by several researchers
from both mathematical and physical points of view 
\cite{APT,manakov,laksh,tsuchida,Kanna1}. 
It was pointed out that vector solitons 
can be used in the construction of logic gate \cite{Kanna2,steiglitz1,steiglitz2}. 
It was also pointed out that the Yang-Baxter map is the key to
understand the mathematical structure of logic gate based on vector
solitons \cite{tsuchida,veselov,APT2}. 

Although there are many works about one-dimensional vector solitons, 
a mathematical work about two-dimensional vector solitons is still
missing. 
For a more complete understanding mathematical structure of 
vector solitons, the study
of two-dimensional vector solitons is very important. 

In this paper, we derive a two-component analogue of two-dimensional 
long wave-short wave resonance interaction (2c-2d-LSRI) equations in
a physical setting. 
We also present the Wronskian solution of the integrable 2c-2d-LSRI equations.
 
\section{Derivation}

Consider the interaction of nonlinear dispersive waves on 3 channels,
e.g. laser beams on some dispersive material. 
Suppose that the dispersion relations of these weakly nonlinear waves are
\[
\omega_i =\omega_i(k_{x,i},k_{y,i}:|A_1|^2, |A_2|^2,|A_3|^2)\,,\quad  {\rm
for} \quad i=1,2,3
\]
where $\omega_i$ and $A_i$ are angular 
frequencies and amplitudes of each channel i, respectively
Suppose that carrier wave is expressed by $\exp(i(k_{x,0}x+k_{y,0}y-\omega_0t))$. 
Taylor expansion around ${\bf k}_0=(k_{x,0}, k_{y,0})$, $\omega_0$ and 
$|A_i|=0$ makes
\begin{eqnarray}
\fl &&\omega_i-\omega_0=\left(\frac{\partial \omega_i}{\partial
 k_{x,i}}\right)_0(k_{x,i}-k_{x,0})+
\left(\frac{\partial \omega_i}{\partial
 k_{y,i}}\right)_0(k_{y,i}-k_{y,0})\nonumber\\
\fl &&\qquad +\frac{1}{2}\left(\frac{\partial^2 \omega_i}{\partial
 k_{x,i}^2}\right)_0(k_{x,i}-k_{x,0})^2
+\frac{1}{2}\left(\frac{\partial^2 \omega_i}{\partial
 k_{y,i}^2}\right)_0(k_{y,i}-k_{y,0})^2
\nonumber\\
\fl && \qquad +\left(\frac{\partial^2 \omega_i}{\partial
 k_{x,i}k_{y,i}}\right)_0(k_{x,i}-k_{x,0})(k_{y,i}-k_{y,0})
\nonumber\\
\fl &&\qquad +\left(\frac{\partial \omega_i}{\partial |A_1|^2} \right)_0|A_1|^2
+\left(\frac{\partial \omega_i}{\partial |A_2|^2} \right)_0|A_2|^2
+\left(\frac{\partial \omega_i}{\partial |A_3|^2} \right)_0|A_3|^2
+\cdots \,,\quad \nonumber \\
\fl && \qquad \qquad  \quad \quad \quad  \quad \quad \quad 
 \quad \quad \quad  \quad \quad \quad  \quad \quad \quad
 \quad \quad \quad  {\rm for}\quad i=1,2,3\,, 
\end{eqnarray}
where the subscript $0$ of $(\quad)_0$ means setting $k_{x,i}=k_{x,0}$,
$k_{y,i}=k_{y,0}$, $\omega_i=\omega_0$ and $|A_i|=0$.
Replacing $\omega_i$, $k_{x,i}$ and $k_{y,i}$ to operators
by the rules $\omega_i-\omega_0 \sim {\rm i}\partial/\partial t$, 
$k_{x,i}-k_{x,0}\sim -{\rm i}\partial /\partial x$, 
$k_{y,i}-k_{y,0}\sim -{\rm i}\partial /\partial y$, 
and applying those equations to $A_i(x,y,t)$, we obtain
\begin{eqnarray}
\fl &&{\rm i}\frac{\partial A_i}{\partial t}+{\rm i} 
\left(\frac{\partial \omega_i}{\partial
 k_{x,i}}\right)_0\frac{\partial A_i}{\partial x}
+{\rm i}\left(\frac{\partial \omega_i}{\partial
 k_{y,i}}\right)_0
\frac{\partial A_i}{\partial y}
\nonumber\\
\fl &&\quad +\frac{1}{2}\left(\frac{\partial^2 \omega_i}{\partial
 k_{x,i}^2}\right)_0
\frac{\partial^2 A_i}{\partial x^2}
+\frac{1}{2}\left(\frac{\partial^2 \omega_i}{\partial
 k_{y,i}^2}\right)_0\frac{\partial^2 A_i}{\partial y^2}
+\left(\frac{\partial^2 \omega_i}{\partial
 k_{x,i}\partial 
k_{y,i}}\right)_0\frac{\partial^2 A_i}{\partial x \partial y}\nonumber\\
\fl &&\quad -\left(\frac{\partial \omega_i}{\partial |A_1|^2} \right)_0|A_1|^2A_i
-\left(\frac{\partial \omega_i}{\partial |A_2|^2} \right)_0|A_2|^2A_i
-\left(\frac{\partial \omega_i}{\partial |A_3|^2}
 \right)_0|A_3|^2A_i=0\,,\nonumber\\
\fl && \quad \quad \quad 
 \quad \quad \quad  \quad \quad \quad 
 \quad \quad \quad  \quad \quad \quad 
 \quad \quad \quad \quad \quad \quad \quad \quad \quad {\rm for}\quad i=1,2,3. 
\end{eqnarray}
By the transformation of coordinate
\[
x'=x-\left(\frac{\partial \omega_3}{\partial
 k_{x,3}}\right)_0t\,, \quad 
y'=y-\left(\frac{\partial \omega_3}{\partial
 k_{y,3}}\right)_0t
\,,\quad t'=t \,,
\]
we obtain
\begin{eqnarray}
\fl &&i\frac{\partial A_1}{\partial t}+i 
v_{x,1}\frac{\partial A_1}{\partial x}
+iv_{y,1}
\frac{\partial A_1}{\partial y}
\nonumber\\
\fl &&\quad +\frac{1}{2}\left(\frac{\partial^2 \omega_1}{\partial
 k_{x,1}^2}\right)_0
\frac{\partial^2 A_1}{\partial x^2}
+\frac{1}{2}\left(\frac{\partial^2 \omega_1}{\partial
 k_{y,1}^2}\right)_0\frac{\partial^2 A_1}{\partial y^2}
+\left(\frac{\partial^2 \omega_1}{\partial
 k_{x,1}\partial k_{y,1}}\right)_0
\frac{\partial^2 A_1}{\partial x \partial y}\nonumber\\
\fl &&\quad 
-\left(\frac{\partial \omega_1}{\partial |A_1|^2} \right)_0|A_1|^2A_1
-\left(\frac{\partial \omega_1}{\partial |A_2|^2} \right)_0|A_2|^2A_1
-\left(\frac{\partial \omega_1}{\partial |A_3|^2} \right)_0|A_3|^2A_1=0\,,
\end{eqnarray}
\begin{eqnarray}
\fl &&i\frac{\partial A_2}{\partial t}+i 
v_{x,2}\frac{\partial A_2}{\partial x}
+iv_{y,2}
\frac{\partial A_2}{\partial y}
\nonumber\\
\fl &&\quad +\frac{1}{2}\left(\frac{\partial^2 \omega_2}{\partial
 k_{x,2}^2}\right)_0
\frac{\partial^2 A_2}{\partial x^2}
+\frac{1}{2}\left(\frac{\partial^2 \omega_2}{\partial
 k_{y,2}^2}\right)_0\frac{\partial^2 A_2}{\partial y^2}
+\left(\frac{\partial^2 \omega_2}{\partial
 k_{x,2}\partial k_{y,2}}\right)_0
\frac{\partial^2 A_2}{\partial x \partial y}\nonumber\\
\fl &&\quad -\left(\frac{\partial \omega_2}{\partial |A_1|^2} \right)_0|A_1|^2A_2
-\left(\frac{\partial \omega_2}{\partial |A_2|^2} \right)_0|A_2|^2A_2
-\left(\frac{\partial \omega_2}{\partial |A_3|^2} \right)_0|A_3|^2A_2=0\,,
\end{eqnarray}
\begin{eqnarray}
\fl &&i\frac{\partial A_3}{\partial t}
+\frac{1}{2}\left(\frac{\partial^2 \omega_3}{\partial
 k_{x,3}^2}\right)_0
\frac{\partial^2 A_3}{\partial x^2}
+\frac{1}{2}\left(\frac{\partial^2 \omega_3}{\partial
 k_{y,3}^2}\right)_0\frac{\partial^2 A_3}{\partial y^2}
+\left(\frac{\partial^2 \omega_3}{\partial
 k_{x,3}\partial k_{y,3}}\right)_0
\frac{\partial^2 A_3}{\partial x \partial y}\nonumber\\
\fl &&\quad 
-\left(\frac{\partial \omega_3}{\partial |A_1|^2} \right)_0|A_1|^2A_3
-\left(\frac{\partial \omega_3}{\partial |A_2|^2} \right)_0|A_2|^2A_3
-\left(\frac{\partial \omega_3}{\partial |A_3|^2} \right)_0|A_3|^2A_3=0\,,
\end{eqnarray}
where
\[
v_{x,1}= \left(\frac{\partial \omega_1}{\partial
 k_{x,1}}\right)_0-\left(\frac{\partial \omega_3}{\partial
 k_{x,3}}\right)_0
\,,\quad 
v_{y,1}= \left(\frac{\partial \omega_1}{\partial
 k_{y,1}}\right)_0-\left(\frac{\partial \omega_3}{\partial
 k_{y,3}}\right)_0\,,
\]
\[
v_{x,2}= \left(\frac{\partial \omega_2}{\partial
 k_{x,2}}\right)_0-\left(\frac{\partial \omega_3}{\partial
 k_{x,3}}\right)_0
\,,\quad 
v_{y,2}= \left(\frac{\partial \omega_2}{\partial
 k_{y,2}}\right)_0-\left(\frac{\partial \omega_3}{\partial
 k_{y,3}}\right)_0\,,
\]
and $x', y', t'$ are replaced by $x, y, t$.
This is a generalization of the vector Nonlinear Schr\"odinger equation
(Manakov system)\cite{manakov,asano}.

Now, consider the situation in which $v_{x,1}$ and $v_{x,2}$ are neglected. 
For notational convenience, we rewrite the above equations as
\begin{eqnarray}
&&i\frac{\partial A_1}{\partial t}
+iv_{y,1}
\frac{\partial A_1}{\partial y}
+\alpha_1
\frac{\partial^2 A_1}{\partial x^2}
+\alpha_2
\frac{\partial^2 A_1}{\partial y^2}
+\alpha_3
\frac{\partial^2 A_1}{\partial x \partial y}\nonumber\\
&&\quad +\alpha_4|A_1|^2A_1
+\alpha_5|A_2|^2A_1
+\alpha_6|A_3|^2A_1=0 \,,\label{cnls1}\\
&&i\frac{\partial A_2}{\partial t}
+iv_{y,2}
\frac{\partial A_2}{\partial y}
+\beta_1
\frac{\partial^2 A_2}{\partial x^2}
+\beta_2\frac{\partial^2 A_2}{\partial y^2}
+\beta_3
\frac{\partial^2 A_2}{\partial x \partial y}\nonumber\\
&&\quad +\beta_4|A_1|^2A_2
+\beta_5|A_2|^2A_2
+\beta_6|A_3|^2A_2=0 \,,\label{cnls2}\\
&&i\frac{\partial A_3}{\partial t}
+\gamma_1
\frac{\partial^2 A_3}{\partial x^2}
+\gamma_2\frac{\partial^2 A_3}{\partial y^2}
+\gamma_3
\frac{\partial^2 A_3}{\partial x \partial y}\nonumber\\
&&\quad +\gamma_4|A_1|^2A_3
+\gamma_5|A_2|^2A_3
+\gamma_6|A_3|^2A_3=0\,. \label{cnls3}
\end{eqnarray}
Assume that the channel 3 is normal dispersion and the channels 1 and 2 are
anomalous dispersion. We study the dark pulses generated in the channel
3:\cite{Kiv} 
\[
\fl A_1=\psi_1\exp({\rm i}\delta_1 t)\,,\quad 
A_2=\psi_2\exp({\rm i}\delta_2 t)\,, \quad 
A_3=(u_0+a(x,y,t))\exp({\rm i}\Gamma t+{\rm i}\phi(x,y,t)))\,,
\]
\[\fl 
\delta_1=-\left(\frac{\partial \omega_1}{\partial |A_3|^2}
\right)_0u_0^2\,,\quad 
\delta_2=-\left(\frac{\partial \omega_2}{\partial |A_3|^2}
\right)_0u_0^2\,,\quad 
\Gamma=-\left(\frac{\partial \omega_3}{\partial |A_3|^2}
\right)_0u_0^2\,, 
\]
where $a$ and $\psi_i (i=1,2)$ are small.
Substituting these into equations (\ref{cnls1})-(\ref{cnls3}), we obtain
\begin{eqnarray}
&&
\frac{\partial a}{\partial t}+\gamma_1 u_0 \frac{\partial^2 \phi}{\partial x^2}
 +\gamma_2 u_0\frac{\partial^2 \phi}{\partial y^2} 
 +\gamma_3 u_0\frac{\partial^2 \phi}{\partial x \partial y}=0\,,\label{p-eq1}\\
&&-u_0 \frac{\partial \phi}{\partial t}
+\gamma_1 \frac{\partial^2 a}{\partial x^2}
+\gamma_2 \frac{\partial^2 a}{\partial y^2}
+\gamma_3 \frac{\partial^2 a}{\partial x \partial y}
\nonumber\\
&&\qquad +\gamma_4 u_0|\psi_1|^2+\gamma_5 u_0|\psi_2|^2
+3\gamma_6 u_0^2a=0\,,\label{p-eq2}\\
&&i\frac{\partial \psi_1}{\partial t}
+iv_{y,1}\frac{\partial \psi_1}{\partial y}
+\alpha_1 \frac{\partial^2 \psi_1}{\partial x^2}
+\alpha_2 \frac{\partial^2 \psi_1}{\partial y^2}
+\alpha_3 \frac{\partial^2 \psi_1}{\partial x \partial y}\nonumber \\
&&\quad +\alpha_4 |\psi_1|^2\psi_1+\alpha_5 |\psi_2|^2\psi_1
+2\alpha_6 u_0 a \psi_1=0 \,,\\
&&i\frac{\partial \psi_2}{\partial t}
+iv_{y,2}\frac{\partial \psi_2}{\partial y}
+\beta_1 \frac{\partial^2 \psi_2}{\partial x^2}
+\beta_2 \frac{\partial^2 \psi_2}{\partial y^2}
+\beta_3 \frac{\partial^2 \psi_2}{\partial x \partial y}\nonumber \\
&&\quad +\beta_4 |\psi_1|^2\psi_2+\beta_5 |\psi_2|^2\psi_2
+2\beta_6 u_0 a \psi_2=0\,. 
\end{eqnarray}
Assume that the $y$-dependency of $\phi$ can be neglected, i.e. we can
neglect $\phi_y$ and $\phi_{yy}$. 
Then eq.(\ref{p-eq1}) reduces to 
\[
 \frac{\partial a}{\partial t}
+\gamma_1 u_0 \frac{\partial^2 \phi}{\partial x^2}=0\,,
\]
i.e., 
\[
\frac{\partial^2 \phi}{\partial x^2}=
-\frac{1}{\gamma_1 u_0 }\frac{\partial a}{\partial t}\,.
\]
Substitute this into eq.(\ref{p-eq2}), we have 
\begin{eqnarray}
&&\frac{\partial^2 a}{\partial t^2}
+3\gamma_1\gamma_6 u_0^2\frac{\partial^2 a}{\partial x^2}
+\gamma_1^2 \frac{\partial^4 a}{\partial x^4}
+\gamma_1 \gamma_2 \frac{\partial^4 a}{\partial x^2\partial y^2}
+\gamma_1 \gamma_3 \frac{\partial^4 a}{\partial x^3 \partial y}
\nonumber\\
&&\qquad +\gamma_1 \frac{\partial^2}{\partial x^2}
(\gamma_4 u_0|\psi_1|^2+\gamma_5 u_0|\psi_2|^2)
=0\,.
\end{eqnarray}
By  
\[
t'=\epsilon t, \quad x'=\epsilon^{1/2}(x+ct), \quad y'=\epsilon y\,, 
\]
($c=3\gamma_1 \gamma_6 u_0^2$)
with $a=\epsilon a_0$, $\psi_1=\epsilon^{3/4}\Phi_1$, 
$\psi_2=\epsilon^{3/4}\Phi_2$ ($\epsilon$ is small), 
we obtain equations of lowest order of
$\epsilon$
\begin{eqnarray}
&&2c\frac{\partial^2 a}{\partial x \partial t}
+\gamma_1 \frac{\partial^2}{\partial x^2}
(\gamma_4 u_0|\psi_1|^2+\gamma_5 u_0|\psi_2|^2)
=0\,,\\
&&i\frac{\partial \psi_1}{\partial t}
+iv_{y,1}\frac{\partial \psi_1}{\partial y}
+\alpha_1 \frac{\partial^2 \psi_1}{\partial x^2}
+2\alpha_6 u_0 a \psi_1=0\,, \\
&&i\frac{\partial \psi_2}{\partial t}
+iv_{y,2}\frac{\partial \psi_2}{\partial y}
+\beta_1 \frac{\partial^2 \psi_2}{\partial x^2}
+2\beta_6 u_0 a \psi_2=0\,.
\end{eqnarray}
Here we again have disregarded the primes and 
have replaced $a_0$, $\Phi_1$ and $\Phi_2$ with $a$, $\psi_1$ and $\psi_2$. 
The first equation leads to 
\begin{equation}
2c\frac{\partial a}{\partial t}
+\gamma_1 \frac{\partial}{\partial x}
(\gamma_4 u_0|\psi_1|^2+\gamma_5 u_0|\psi_2|^2)
=0\,.
\end{equation}
This system is 
nothing but the 2-component analogue of 2-dimensional analogue of the 
long wave-short wave
resonance interaction (2c-2d-LSRI)
equations\cite{benney,yajima-oikawa,oikawa,melnikov}. 
Note that the special case of 
coefficients ($v_{y,1}=v_{y,2}, \alpha_1=\beta_1, \alpha_6=\beta_6, \gamma_4=\gamma_5$) is integrable. In the one-component
integrable case, several solutions have been presented 
\cite{oikawa,chow,radha,yurova}. 
In Ref.\cite{berkela}, some solutions of matrix generalization
were discussed. In the next section, we will present the determinant formula of
$N$-soliton solution for the integrable 2c-2d-LSRI equations. 

\section{Soliton Solutions} 
We study the soliton solutions of an integrable case of the 
2c-2d-LSRI equations
\begin{eqnarray}
&&i(S_t^{(1)}+S_y^{(1)})-S_{xx}^{(1)}+LS^{(1)}=0\,,\\
&&i(S_t^{(2)}+S_y^{(2)})-S_{xx}^{(2)}+L S^{(2)}=0\,,\\
&&L_t=2(|S^{(1)}|^2)_x+2(|S^{(2)}|^2)_x\,.
\end{eqnarray}
By the dependent variable transformation
\begin{equation}
S^{(1)}=\frac{G}{F}\,,\quad 
S^{(2)}=\frac{H}{F}\,,\quad 
L=-2\frac{\partial^2}{\partial x^2}\log F\,,
\end{equation}
we have three bilinear equations
\begin{eqnarray}
&&[i(D_t+D_y)-D_x^2]G \cdot F=0\,,\label{bilinear1}\\
&&[i(D_t+D_y)-D_x^2]H \cdot F=0\,,\label{bilinear2}\\
&&(D_tD_x-2c)F\cdot F+2GG^*+2HH^*=0\,.\label{bilinear3}
\end{eqnarray}
Here we set $c=0$ which means we consider the bright-type soliton solutions. 
These bilinear equations belong to the three-component KP hierarchy
\cite{DJKM1,DJKM2,kac,HirotaBook}. 
We can construct soliton solutions in which the number of solitons in 
channel $i$ is $N_i$ ($i=1,2,3$). 
We call this solution $(N_1,N_2,N_3)$-soliton solution. 

{\bf Wronskian form of the bright type $(N,M,N+M)$-soliton solutions}

Let 
\small
\begin{eqnarray*}
\fl &&\tau_{nm}^{NM}=\\
\fl &&\left|\begin{array}{cccccccccccc}
\varphi_1 &\varphi_1^{(1)} &\cdots &\varphi_1^{(N+M-1+n+m)}
&\psi_1 &\psi_1^{(1)} &\cdots &\psi_1^{(N-1-n)}
&0 &0 &\cdots &0
\cr
\varphi_2 &\varphi_2^{(1)} &\cdots &\varphi_2^{(N+M-1+n+m)}
&\psi_2 &\psi_2^{(1)} &\cdots &\psi_2^{(N-1-n)}
&0 &0 &\cdots &0
\cr
\vdots &\vdots &&\vdots
&\vdots &\vdots &&\vdots
&\vdots &\vdots &&\vdots
\cr
\varphi_{2N} &\varphi_{2N}^{(1)} &\cdots &\varphi_{2N}^{(N+M-1+n+m)}
&\psi_{2N} &\psi_{2N}^{(1)} &\cdots &\psi_{2N}^{(N-1-n)}
&0 &0 &\cdots &0
\cr
\phi_1 &\phi_1^{(1)} &\cdots &\phi_1^{(N+M-1+n+m)}
&0 &0 &\cdots &0
&\chi_1 &\chi_1^{(1)} &\cdots &\chi_1^{(M-1-m)}
\cr
\phi_2 &\phi_2^{(1)} &\cdots &\phi_2^{(N+M-1+n+m)}
&0 &0 &\cdots &0
&\chi_2 &\chi_2^{(1)} &\cdots &\chi_2^{(M-1-m)}
\cr
\vdots &\vdots &&\vdots
&\vdots &\vdots &&\vdots
&\vdots &\vdots &&\vdots
\cr
\phi_{2M} &\phi_{2M}^{(1)} &\cdots &\phi_{2M}^{(N+M-1+n+m)}
&0 &0 &\cdots &0
&\chi_{2M} &\chi_{2M}^{(1)} &\cdots &\chi_{2M}^{(M-1-m)}
\cr
\end{array}\right|
\end{eqnarray*}
\normalsize
\begin{eqnarray*}
\fl &&\varphi_i=e^{\xi_i}\,,
\qquad
\xi_i=p_ix_1+p_i^2x_2\,,
\qquad
\hbox{for \, $i=1,2,\cdots,N$}
\\
\fl &&\varphi_{N+i}=e^{-\xi_i^*}\,,
\qquad
-\xi_i^*=-p_i^*x_1+(-p_i^*)^2x_2\,,
\qquad
\hbox{for \, $i=1,2,\cdots,N$}
\\
\fl &&\phi_i=e^{\theta_i}\,,
\qquad
\theta_i=s_ix_1+s_i^2x_2\,,
\qquad
\hbox{for \, $i=1,2,\cdots,M$}
\\
\fl &&\phi_{M+i}=e^{-\theta_i^*}\,,
\qquad
-\theta_i^*=-s_i^*x_1+(-s_i^*)^2x_2\,,
\qquad
\hbox{for \, $i=1,2,\cdots,M$}
\\
\fl &&\psi_i=a_ie^{\eta_i}\,,
\qquad
\eta_i=q_iy_1+\eta_{i0}\,,
\qquad
\hbox{for \, $i=1,2,\cdots,N$}
\\
\fl &&\psi_{N+i}=a_{N+i}e^{-\eta_i^*}\,,
\qquad
-\eta_i^*=-q_i^*y_1-\eta_{i0}^*\,,
\qquad
\hbox{for \, $i=1,2,\cdots,N$}
\\
\fl &&\chi_i=b_ie^{\zeta_i}\,,
\qquad
\zeta_i=r_iz_1+\zeta_{i0}\,,
\qquad
\hbox{for \, $i=1,2,\cdots,M$}
\\
\fl &&\chi_{M+i}=b_{M+i}e^{-\zeta_i^*}\,,
\qquad
-\zeta_i^*=-r_i^*z_1-\zeta_{i0}^*\,,
\qquad
\hbox{for \, $i=1,2,\cdots,M$}
\end{eqnarray*}
\begin{eqnarray*}
\fl &&a_i=
\left(\prod_{k=1 \atop k\ne i}^N\frac{p_k-p_i}{q_k-q_i}\right)
\left(\prod_{l=1}^M(s_l-p_i)\right)\,,
\qquad
\hbox{for \, $i=1,2,\cdots,N$}
\\
\fl &&a_{N+i}=\epsilon_i
\left(\prod_{k=1}^N\frac{p_k+p_i^*}{q_k+q_i^*}\right)
\left(\prod_{l=1}^M(s_l+p_i^*)\right)\,,
\qquad
\hbox{for \, $i=1,2,\cdots,N$}
\\
\fl &&b_i=
\left(\prod_{k=1}^N(p_k-s_i)\right)
\left(\prod_{l=1 \atop l\ne i}^M\frac{s_l-s_i}{r_l-r_i}\right)\,,
\qquad
\hbox{for \, $i=1,2,\cdots,M$}
\\
\fl &&b_{M+i}=\delta_i
\left(\prod_{k=1}^N(p_k+s_i^*)\right)
\left(\prod_{l=1}^M\frac{s_l+s_i^*}{r_l+r_i^*}\right)\,,
\qquad
\hbox{for \, $i=1,2,\cdots,M$}
\end{eqnarray*}
$$
\epsilon_i=\pm1\,,
\qquad
\delta_i=\pm1\,,
$$
where ${}^*$ means complex conjugate and
$p_i$, $q_i$ ($1\le i\le N$) and $s_i$, $r_i$ ($1\le i\le M$)
are complex wave numbers, and
$\eta_{i0}$ ($1\le i\le N$) and $\zeta_{i0}$ ($1\le i\le M$)
are complex phase parameters.
In order to obtain regular solutions, we have to choose
appropriate sign for $\epsilon_i$ and $\delta_i$,
which depend on parameters $p_i,q_i,r_i,s_i$.
We take
$$
x_1=x\,,
\qquad
x_2=-iy\,,
\qquad
y_1=y-t\,,
\qquad
z_1=y-t\,,
$$
where $x$, $y$ and $t$ are real\,,
(i.e. $x_1$, $y_1$ and $z_1$ are real and $x_2$ is pure imaginary).

Let
$$
f=\tau_{00}\,,
\qquad
g=\tau_{10}\,,
\qquad
\bar g=\tau_{-1,0}\,,
\qquad
h=\tau_{01}\,,
\qquad
\bar h=\tau_{0,-1}\,.
$$
These tau-functions satisfy the condition 
$$
\left(\frac{g}{f}\right)^*=\frac{\bar g}{f}\,,
\qquad
\left(\frac{h}{f}\right)^*=\frac{\bar h}{f}
\,,$$
$$
f \mathcal{G}  \hbox{ : real}\,.
$$
where $\mathcal{G}$ is an exponential factor which is a gauge function
(see Appendix).
Let $F=f\mathcal{G}$, 
$G=g\mathcal{G}$,
$G^*=\bar{g}\mathcal{G}$, 
$H=h\mathcal{G}$, 
$H^*=\bar{h}\mathcal{G}$. 
The functions $F$, $G$ and $H$ satisfy the bilinear equations 
(\ref{bilinear1})-(\ref{bilinear3}) and reality of $F$ and complex conjugacy of
$G$ and $H$. 
The function $L=-2\frac{\partial^2}{\partial x^2}\log F$ 
represents $N+M$-soliton solution, 
$S_1=G/F$ represents $N$-soliton solution, and 
$S_2=H/F$ represents $M$-soliton solution.

{\bf $(1,1,2)$-soliton solution}\\
The $\tau$-functions of $(1,1,2)$-soliton solution are the following:
\begin{eqnarray*}
\fl &&f=
\left|\begin{array}{cccc}
\varphi_1 &\varphi_1^{(1)} 
&\psi_1 &0
\cr
\varphi_2 &\varphi_2^{(1)} 
&\psi_2 &0
\cr
\phi_1 &\phi_1^{(1)} &0 
&\chi_1 
\cr
\phi_2 &\phi_2^{(1)} &0
&\chi_2 
\cr
\end{array}\right|\,,\\
&& g=
\left|\begin{array}{cccc}
\varphi_1 &\varphi_1^{(1)} 
&\varphi_1^{(2)} &0
\cr
\varphi_2 &\varphi_2^{(1)} 
&\varphi_2^{(2)} &0
\cr
\phi_1 &\phi_1^{(1)} & \phi_1^{(2)}
&\chi_1 
\cr
\phi_2 &\phi_2^{(1)} &\phi_2^{(2)}
&\chi_2 
\cr
\end{array}\right|
\,,\quad \bar{g}=\left|\begin{array}{cccc}
\varphi_1  
&\psi_1 &\psi_1^{(1)} &0
\cr
\varphi_2  
&\psi_2 &\psi_2^{(1)}  &0
\cr
\phi_1 & 0 &0 
&\chi_1 
\cr
\phi_2 & 0  &0
&\chi_2 
\cr
\end{array}\right|\,,
\end{eqnarray*}
\begin{eqnarray*}
&&
h=
\left|\begin{array}{cccc}
\varphi_1 &\varphi_1^{(1)} 
&\varphi_1^{(2)} & \psi_1
\cr
\varphi_2 &\varphi_2^{(1)} 
& \varphi_2^{(2)}  &  \psi_2 
\cr
\phi_1 &\phi_1^{(1)} & \phi_1^{(2)} 
& 0
\cr
\phi_2 &\phi_2^{(1)} &\phi_2^{(2)}
&0 
\cr
\end{array}\right|
\,,\quad \bar{h}=
\left|\begin{array}{cccc}
\varphi_1 &\psi_1 &0 &0
\cr
\varphi_2 &\psi_2 &0 & 0
\cr
\phi_1  &0 
&\chi_1 &\chi_1^{(1)}
\cr
\phi_2 &0
&\chi_2 &\chi_2^{(1)}
\cr
\end{array}\right|\,,
\end{eqnarray*}
\begin{eqnarray*}
\fl &&\varphi_1=e^{\xi_1}\,,
\qquad
\xi_1=p_1x_1+p_1^2x_2\,,
\qquad \varphi_{2}=e^{-\xi_1^*}\,,
\qquad
-\xi_1^*=-p_1^*x_1+(-p_1^*)^2x_2\,,
\\
\fl &&
\phi_1=e^{\theta_1}\,,
\qquad
\theta_1=s_1x_1+s_1^2x_2\,,
\qquad
\phi_{2}=e^{-\theta_1^*}\,,
\qquad
-\theta_1^*=-s_1^*x_1+(-s_1^*)^2x_2\,,
\\
\fl &&\psi_1=a_1e^{\eta_1}\,,
\qquad
\eta_1=q_1y_1+\eta_{10}\,,
\qquad
\psi_{2}=a_{2}e^{-\eta_1^*}\,,
\qquad
-\eta_1^*=-q_1^*y_1-\eta_{10}^*\,,
\\
\fl &&\chi_1=b_1e^{\zeta_1}\,,
\qquad
\zeta_1=r_1z_1+\zeta_{10}\,,
\qquad
\chi_{2}=b_{2}e^{-\zeta_1^*}\,,
\qquad
-\zeta_1^*=-r_1^*z_1-\zeta_{10}^*\,,\\
\fl &&a_1=s_1-p_1\,,\quad  b_1=p_1-s_1\,,\quad 
a_2=\epsilon (s_1+p_1^*)\frac{p_1+p_1^*}{q_1+q_1^*}\,,
\quad 
b_2=\delta (s_1^*+p_1)\frac{s_1+s_1^*}{r_1+r_1^*}\,,\\
\fl &&\epsilon=\pm 1, \quad \delta=\pm 1\,,
\end{eqnarray*}
and $x_1=x$, $x_2=-iy$, $y_1=y-t$, $z_1=y-t$. 

\begin{figure}[t!]
\centerline{
\includegraphics[scale=0.4]{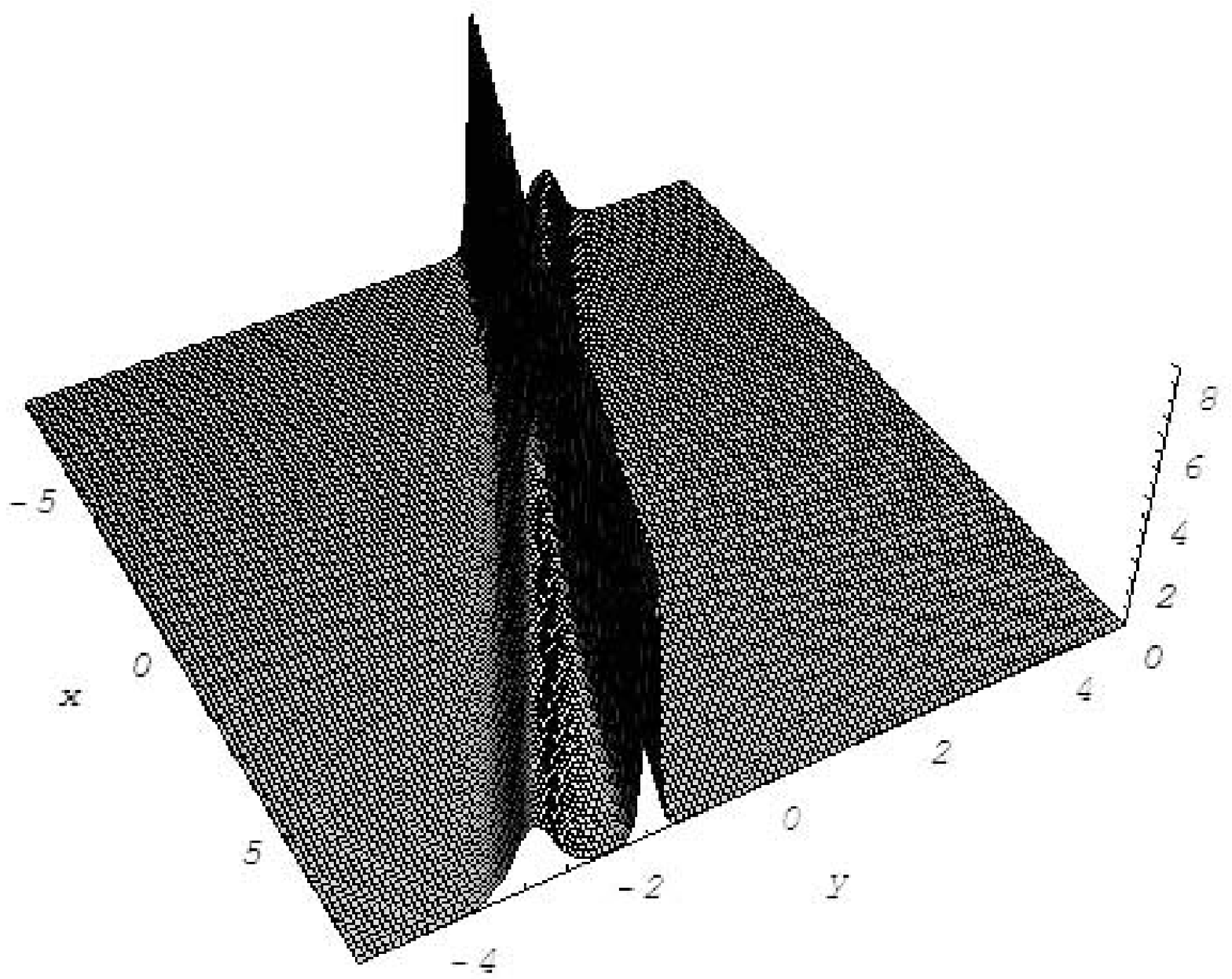}
\includegraphics[scale=0.4]{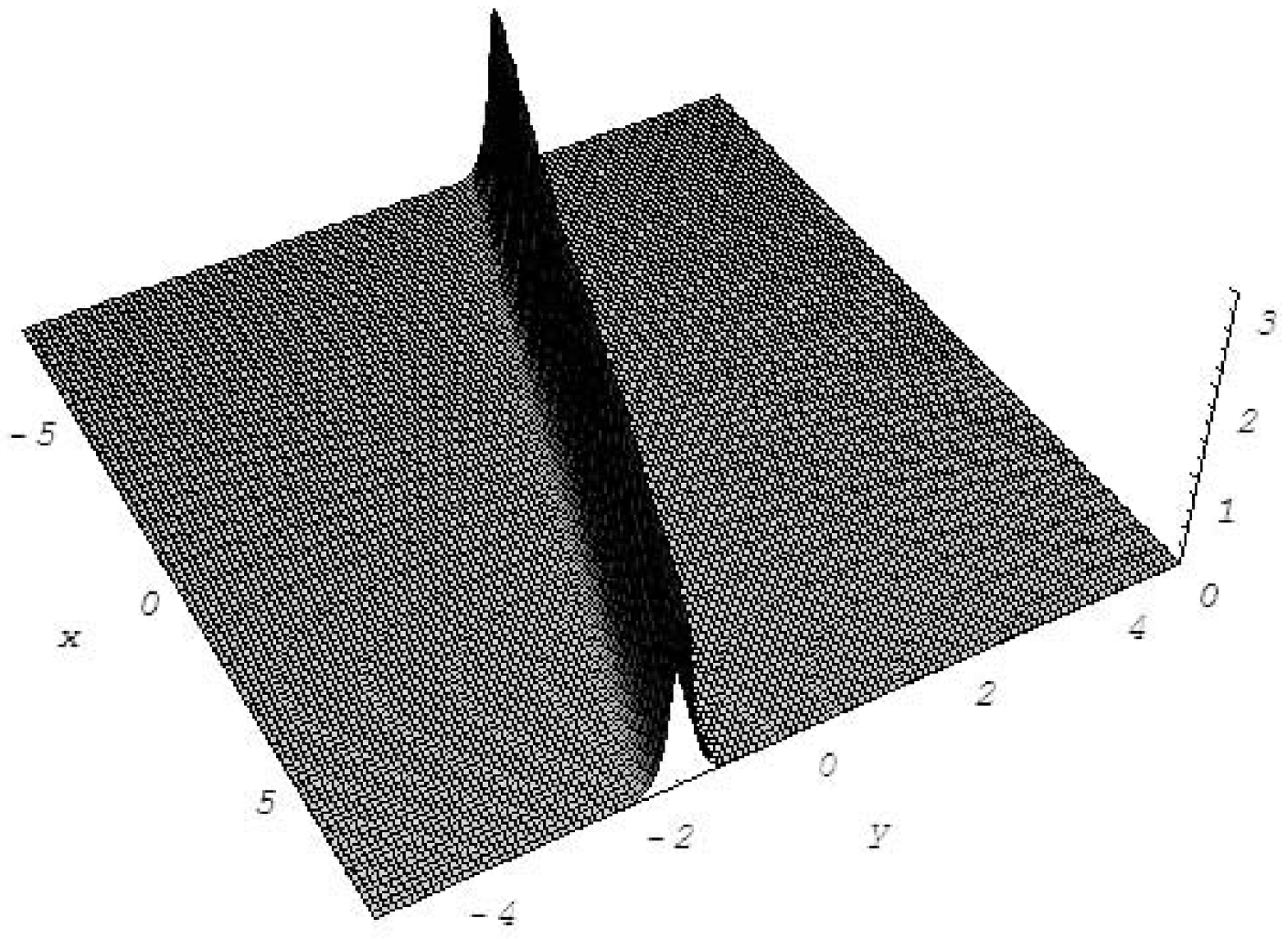}
}
\centerline{
\includegraphics[scale=0.43]{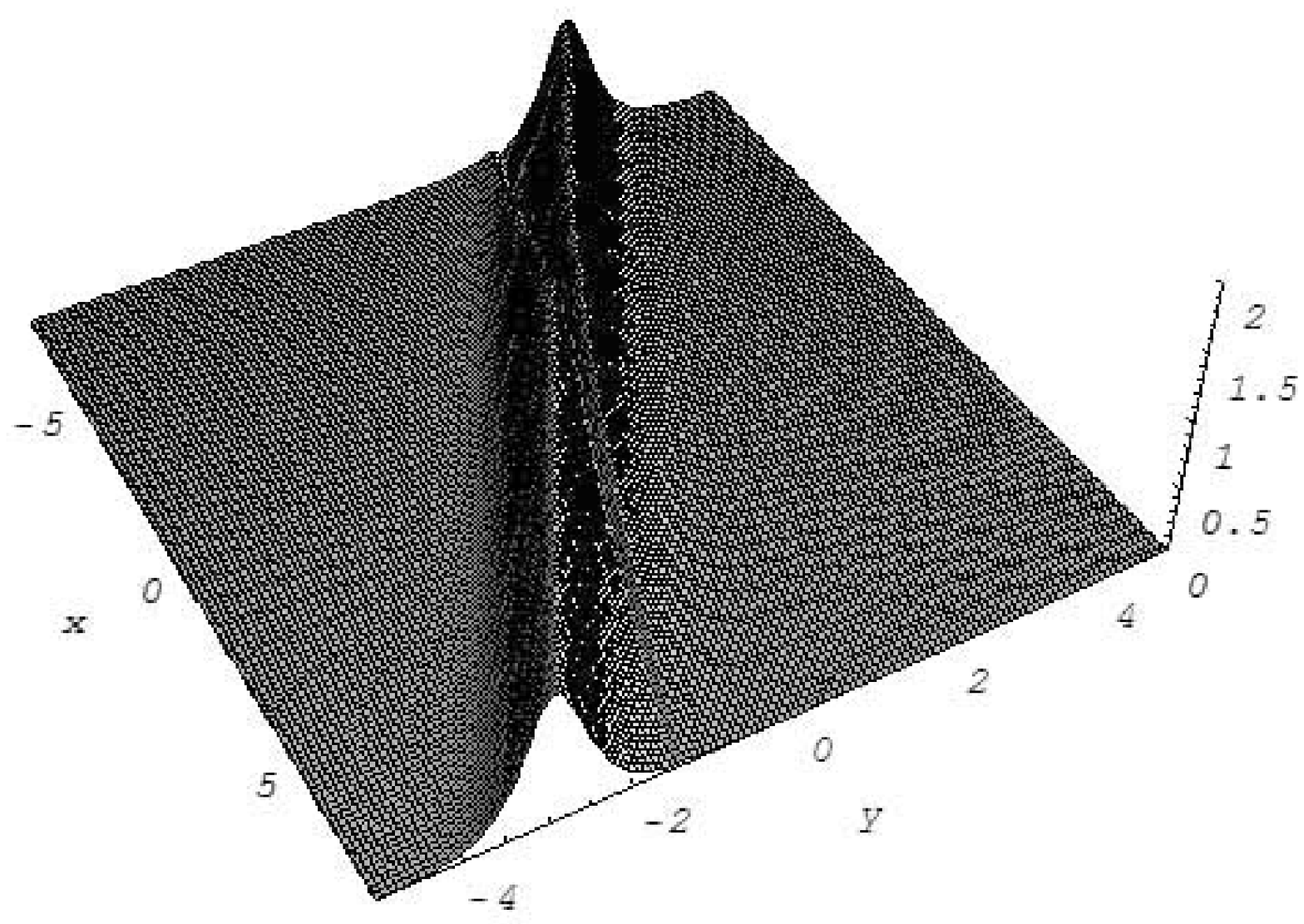}
}
\caption{$(1,1,2)$-soliton solution. 
$p_1=2+2i$, $s_1=-1+i$, $q_1=-2+i$, $r_1=1+i$, $\epsilon=\delta=1$.
The top left graph is $-L$, the top right graph is
 $S^{(1)}$ and the bottom graph is $S^{(2)}$ at $t=0$.}
\label{fig1}
\end{figure}

Figure \ref{fig1} shows the $(1,1,2)$-soliton. 
In the field $S_1$ and $S_2$, there is the single soliton.  
However, each solitons in $S_1$ and $S_2$
interact through the field $L$. So the behavior of solitons looks 
like two-soliton solution, i.e., we see the phase shift in the
region of the interaction (see the bottom graph in Fig.\ref{fig1}). 

{\bf $(2,2,4)$-soliton solution}\\
The $\tau$-functions of $(2,2,4)$-soliton solution are the following:
\begin{eqnarray*}
&&f=
\left|\begin{array}{cccccccc}
\varphi_1 &\varphi_1^{(1)}&\varphi_1^{(2)}&\varphi_1^{(3)} 
&\psi_1 &\psi_1^{(1)} &0 &0
\cr
\varphi_2 &\varphi_2^{(1)} &\varphi_2^{(2)} &\varphi_2^{(3)}
&\psi_2 &\psi_2^{(1)} & 0 &0
\cr
\varphi_3 &\varphi_3^{(1)} &\varphi_3^{(2)} &\varphi_3^{(3)}
&\psi_3 &\psi_3^{(1)} & 0 &0
\cr
\varphi_4 &\varphi_4^{(1)} &\varphi_4^{(2)} &\varphi_4^{(3)}
&\psi_4 &\psi_4^{(1)} &0 &0
\cr
\phi_1 &\phi_1^{(1)} &\phi_1^{(2)} &\phi_1^{(3)} &0 &0 
&\chi_1 &\chi_1^{(1)} 
\cr
\phi_2 &\phi_2^{(1)} & \phi_2^{(2)} &\phi_2^{(3)} &0 &0
&\chi_2 &\chi_2^{(1)} 
\cr
\phi_3 &\phi_3^{(1)} & \phi_3^{(2)} &\phi_3^{(3)} &0 &0
&\chi_3 &\chi_3^{(1)} 
\cr
\phi_4 &\phi_4^{(1)} & \phi_4^{(2)} &\phi_4^{(3)} &0 &0
&\chi_4 &\chi_4^{(1)} 
\cr
\end{array}\right|
\end{eqnarray*}
\begin{eqnarray*}
&&g=
\left|\begin{array}{cccccccc}
\varphi_1 &\varphi_1^{(1)} 
&\varphi_1^{(2)} & \varphi_1^{(3)} & \varphi_1^{(4)} & \psi_1 & 0 & 0
\cr
\varphi_2 &\varphi_2^{(1)} 
&\varphi_2^{(2)} & \varphi_2^{(3)} & \varphi_2^{(4)} & \psi_2 & 0 & 0
\cr
\varphi_3 &\varphi_3^{(1)} 
&\varphi_3^{(2)} & \varphi_3^{(3)} & \varphi_3^{(4)} & \psi_3 & 0 & 0
\cr
\varphi_4 &\varphi_4^{(1)} 
&\varphi_4^{(2)} & \varphi_4^{(3)} & \varphi_4^{(4)} & \psi_4 & 0 & 0
\cr
\phi_1 &\phi_1^{(1)} & \phi_1^{(2)} & \phi_1^{(3)} & \phi_1^{(4)}
&0 &\chi_1 &\chi_1^{(1)}  
\cr
\phi_2 &\phi_2^{(1)} & \phi_2^{(2)} & \phi_2^{(3)} & \phi_2^{(4)}
&0 &\chi_2 &\chi_2^{(1)}  
\cr
\phi_3 &\phi_3^{(1)} & \phi_3^{(2)} & \phi_3^{(3)} & \phi_3^{(4)}
&0 &\chi_3 &\chi_3^{(1)}
\cr
\phi_4 &\phi_4^{(1)} & \phi_4^{(2)} & \phi_4^{(3)} & \phi_4^{(4)}
&0 &\chi_4 &\chi_4^{(1)}
\end{array}\right|
\end{eqnarray*}
\begin{eqnarray*}
&&
\bar{g}=
\left|\begin{array}{cccccccc}
\varphi_1 &\varphi_1^{(1)}&\varphi_1^{(2)}
&\psi_1 &\psi_1^{(1)} & \psi_1^{(2)} & 0 &0
\cr
\varphi_2 &\varphi_2^{(1)} &\varphi_2^{(2)} &
\psi_2 &\psi_2^{(1)} &\psi_2^{(2)} & 0 &0
\cr
\varphi_3 &\varphi_3^{(1)} &\varphi_3^{(2)} 
&\psi_3 &\psi_3^{(1)} &\psi_3^{(2)}& 0 &0
\cr
\varphi_4 &\varphi_4^{(1)} &\varphi_4^{(2)} 
&\psi_4 &\psi_4^{(1)} &\psi_4^{(2)} &0 &0
\cr
\phi_1 &\phi_1^{(1)} &\phi_1^{(2)} &0 &0 &0 
&\chi_1 &\chi_1^{(1)} 
\cr
\phi_2 &\phi_2^{(1)} & \phi_2^{(2)} &0 &0 &0
&\chi_2 &\chi_2^{(1)} 
\cr
\phi_3 &\phi_3^{(1)} & \phi_3^{(2)} & 0 &0 &0
&\chi_3 &\chi_3^{(1)} 
\cr
\phi_4 &\phi_4^{(1)} & \phi_4^{(2)} & 0 &0 &0
&\chi_4 &\chi_4^{(1)} 
\cr
\end{array}\right|
\end{eqnarray*}
\begin{eqnarray*}
&&
h=
\left|\begin{array}{cccccccc}
\varphi_1 &\varphi_1^{(1)} 
&\varphi_1^{(2)} & \varphi_1^{(3)} & \varphi_1^{(4)} & \psi_1 &
\psi_1^{(1)} 
& 0
\cr
\varphi_2 &\varphi_2^{(1)} 
&\varphi_2^{(2)} & \varphi_2^{(3)} & \varphi_2^{(4)} & \psi_2 &
\psi_2^{(1)} 
& 0
\cr
\varphi_3 &\varphi_3^{(1)} 
&\varphi_3^{(2)} & \varphi_3^{(3)} & \varphi_3^{(4)} & \psi_3 &
\psi_3^{(1)} 
& 0
\cr
\varphi_4 &\varphi_4^{(1)} 
&\varphi_4^{(2)} & \varphi_4^{(3)} & \varphi_4^{(4)} & \psi_4 
& \psi_4^{(1)} & 0
\cr
\phi_1 &\phi_1^{(1)} & \phi_1^{(2)} & \phi_1^{(3)} & \phi_1^{(4)}
&0 & 0 &\chi_1   
\cr
\phi_2 &\phi_2^{(1)} & \phi_2^{(2)} & \phi_2^{(3)} & \phi_2^{(4)}
&0 & 0 &\chi_2
\cr
\phi_3 &\phi_3^{(1)} & \phi_3^{(2)} & \phi_3^{(3)} & \phi_3^{(4)}
&0 & 0 &\chi_3 
\cr
\phi_4 &\phi_4^{(1)} & \phi_4^{(2)} & \phi_4^{(3)} & \phi_4^{(4)}
&0 & 0 &\chi_4 
\end{array}\right|
\end{eqnarray*}
\begin{eqnarray*}
&&
\bar{h}=
\left|\begin{array}{cccccccc}
\varphi_1 &\varphi_1^{(1)}&\varphi_1^{(2)}
&\psi_1 &\psi_1^{(1)} & 0 & 0 &0
\cr
\varphi_2 &\varphi_2^{(1)} &\varphi_2^{(2)} &
\psi_2 &\psi_2^{(1)} & 0 & 0 &0
\cr
\varphi_3 &\varphi_3^{(1)} &\varphi_3^{(2)} 
&\psi_3 &\psi_3^{(1)} & 0 & 0 &0
\cr
\varphi_4 &\varphi_4^{(1)} &\varphi_4^{(2)} 
&\psi_4 &\psi_4^{(1)} & 0 &0 &0
\cr
\phi_1 &\phi_1^{(1)} &\phi_1^{(2)} &0 &0 
&\chi_1 &\chi_1^{(1)} &\chi_1^{(2)} 
\cr
\phi_2 &\phi_2^{(1)} & \phi_2^{(2)} &0 &0
&\chi_2 &\chi_2^{(1)} &\chi_2^{(2)} 
\cr
\phi_3 &\phi_3^{(1)} & \phi_3^{(2)} &0 &0
&\chi_3 &\chi_3^{(1)} &\chi_3^{(2)} 
\cr
\phi_4 &\phi_4^{(1)} & \phi_4^{(2)} &0 &0
&\chi_4 &\chi_4^{(1)} &\chi_4^{(2)}
\cr
\end{array}\right|
\end{eqnarray*}
\begin{eqnarray*}
\fl &&\varphi_1=e^{\xi_1}\,,
\qquad
\xi_1=p_1x_1+p_1^2x_2\,,
\qquad \varphi_2=e^{\xi_2}\,,
\qquad
\xi_2=p_2x_1+p_2^2x_2\,,
\\
\fl &&
\varphi_{3}=e^{-\xi_1^*}\,,
\qquad
-\xi_1^*=-p_1^*x_1+(-p_1^*)^2x_2\,,
\qquad \varphi_4=e^{-\xi_2^*}\,,
\qquad
-\xi_2^*=-p_2^*x_1+(-p_2^*)^2x_2\,,
\\
\fl &&
\phi_1=e^{\theta_1}\,,
\qquad
\theta_1=s_1x_1+s_1^2x_2\,,
\qquad \phi_2=e^{\theta_2}\,,
\qquad
\theta_2=s_2x_1+s_2^2x_2\,,
\\
\fl &&
\phi_{3}=e^{-\theta_1^*}\,,
\qquad
-\theta_1^*=-s_1^*x_1+(-s_1^*)^2x_2\,,
\qquad
\phi_{4}=e^{-\theta_2^*}\,,
\qquad
-\theta_2^*=-s_2^*x_1+(-s_2^*)^2x_2\,,
\\
\fl &&\psi_1=a_1e^{\eta_1}\,,
\qquad
\eta_1=q_1y_1+\eta_{10}\,,\qquad
\psi_2=a_2e^{\eta_2}\,,
\qquad
\eta_2=q_2y_1+\eta_{20}\,,
\\
\fl &&
\psi_{3}=a_{3}e^{-\eta_1^*}\,,
\qquad
-\eta_1^*=-q_1^*y_1-\eta_{10}^*\,,
\qquad
\psi_{4}=a_{4}e^{-\eta_2^*}\,,
\qquad
-\eta_2^*=-q_2^*y_1-\eta_{20}^*\,,
\\
\fl &&\chi_1=b_1e^{\zeta_1}\,,
\qquad
\zeta_1=r_1z_1+\zeta_{10}\,,
\qquad
\chi_2=b_2e^{\zeta_2}\,,
\qquad
\zeta_2=r_2z_1+\zeta_{20}\,,\\
\fl &&
\chi_{3}=b_{3}e^{-\zeta_1^*}\,,
\qquad
-\zeta_1^*=-r_1^*z_1-\zeta_{10}^*\,,
\qquad
\chi_{4}=b_{4}e^{-\zeta_2^*}\,,
\qquad
-\zeta_2^*=-r_2^*z_1-\zeta_{20}^*\,,
\end{eqnarray*}
\begin{eqnarray*}
\fl &&a_1=\frac{p_2-p_1}{q_{2}-q_1}(s_1-p_1)(s_2-p_1)\,,\quad  
b_1=\frac{s_2-s_1}{r_{2}-r_1}(p_1-s_1)(p_2-s_1)\,,\\
\fl &&
a_2=\frac{p_1-p_2}{q_{1}-q_2}(s_1-p_2)(s_2-p_2)\,,\quad  
b_2=\frac{s_1-s_2}{r_{1}-r_2}(p_1-s_2)(p_2-s_2)\,,\\
\fl && 
a_3=\epsilon_1 \frac{p_1+p_1^*}{q_1+q_1^*}
\frac{p_2+p_1^*}{q_2+q_1^*}(s_1+p_1^*)(s_2+p_1^*)\,,\quad
b_3=\delta_1
\frac{s_1+s_1^*}{r_1+r_1^*}\frac{s_2+s_1^*}{r_2+r_1^*}
(s_1^*+p_1)(s_1^*+p_2)\,,\\
\fl && 
a_4=\epsilon_2 \frac{p_1+p_2^*}{q_1+q_2^*}
\frac{p_2+p_2^*}{q_2+q_2^*}(s_1+p_2^*)(s_2+p_2^*)\,,\quad
b_4=\delta_2 \frac{s_1+s_2^*}{r_1+r_2^*}\frac{s_2+s_2^*}{r_2+r_2^*}
(s_2^*+p_1)(s_2^*+p_2)\,,\\
\fl &&\epsilon_i=\pm 1, \quad \delta_i=\pm 1\,\, (i=1,2)\,,
\end{eqnarray*}
and $x_1=x$, $x_2=-iy$, $y_1=y-t$, $z_1=y-t$. 

\begin{figure}[t!]
\centerline{
\includegraphics[scale=0.4]{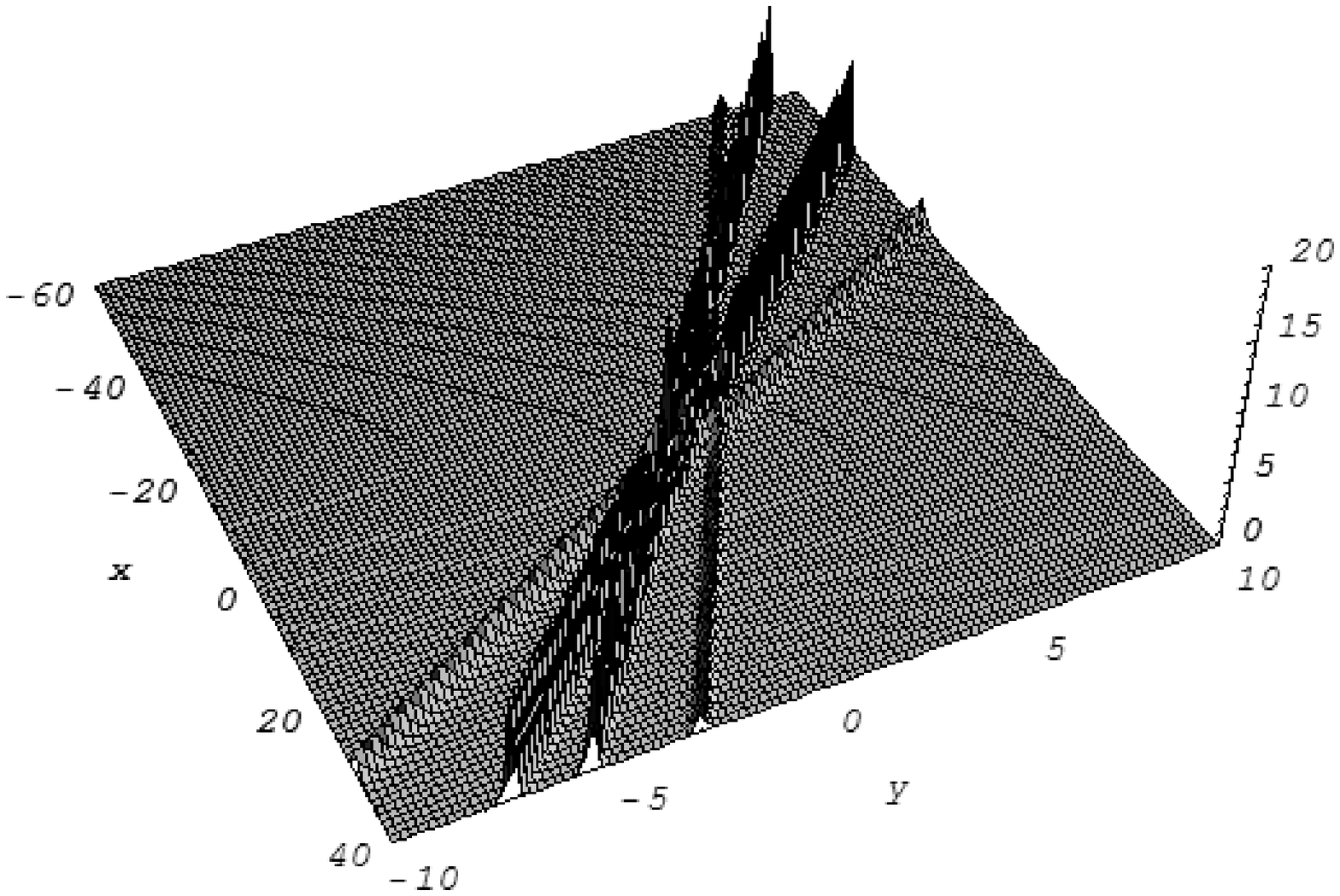}
\includegraphics[scale=0.35]{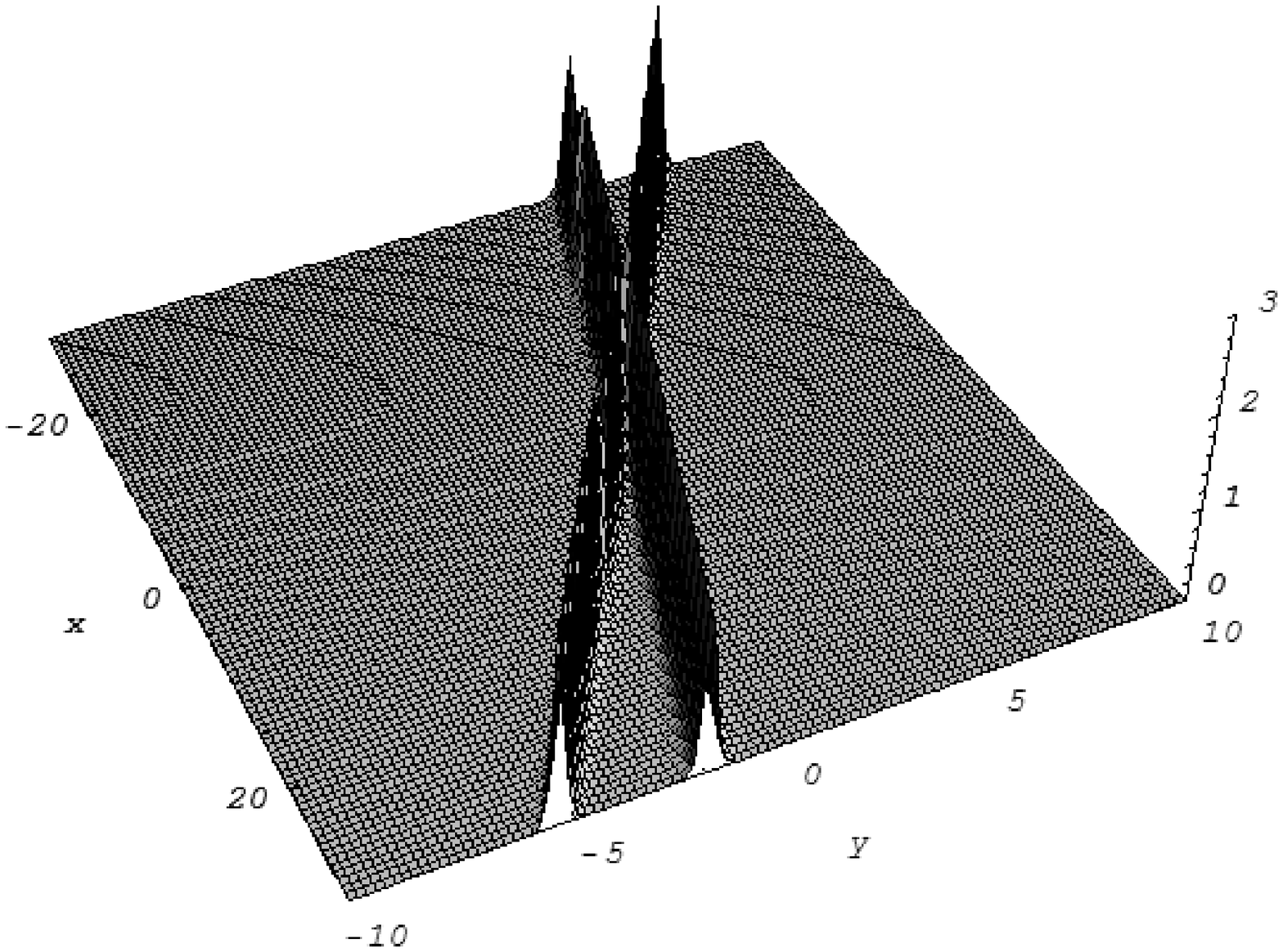}
}
\centerline{
\includegraphics[scale=0.35]{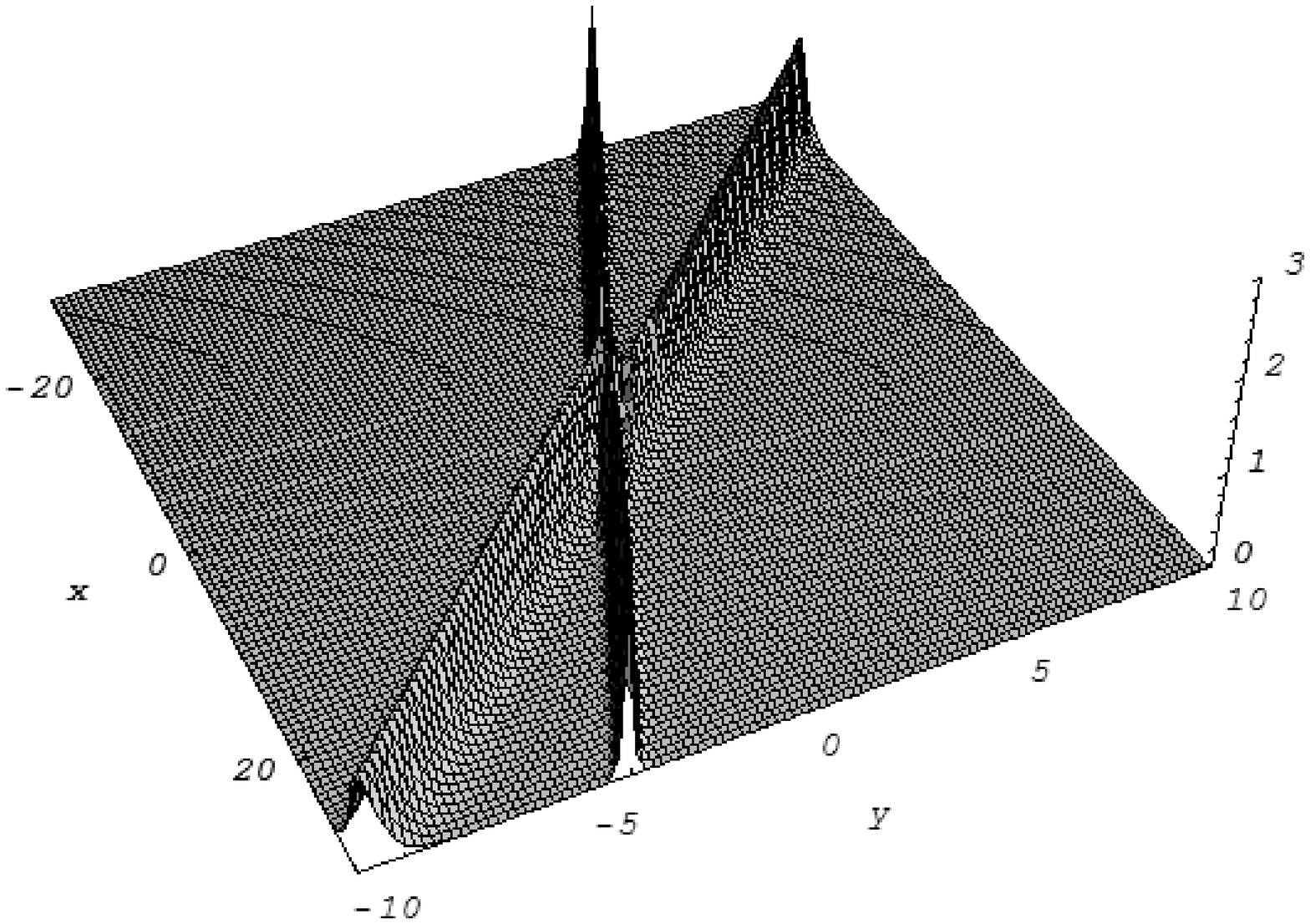}}
\caption{$(2,2,4)$-soliton solution. 
$p_1=2+2i$, $p_2=1+3i$, 
$p_3=2-2i$, $p_4=1-3i$, 
$s_1=-1+i$, $s_2=-2+3i$,
$s_3=-1-i$, $s_4=-2-3i$, 
$q_1=-2+i$, $q_2=-3+2i$, 
$q_3=-2-i$, $q_4=-3-2i$, 
$r_1=1+i$, $r_2=1.5+i$, 
$r_3=1-i$, $r_4=1.5-i$, 
$\epsilon_1=\epsilon_2=\delta_1=\delta_2=1$.
The top left graph is $-L$, the top right graph is
 $S^{(1)}$ and the bottom graph is $S^{(2)}$ at $t=3$.}
\label{fig2}
\end{figure}

\begin{figure}[t!]
\centerline{
\includegraphics[scale=0.35]{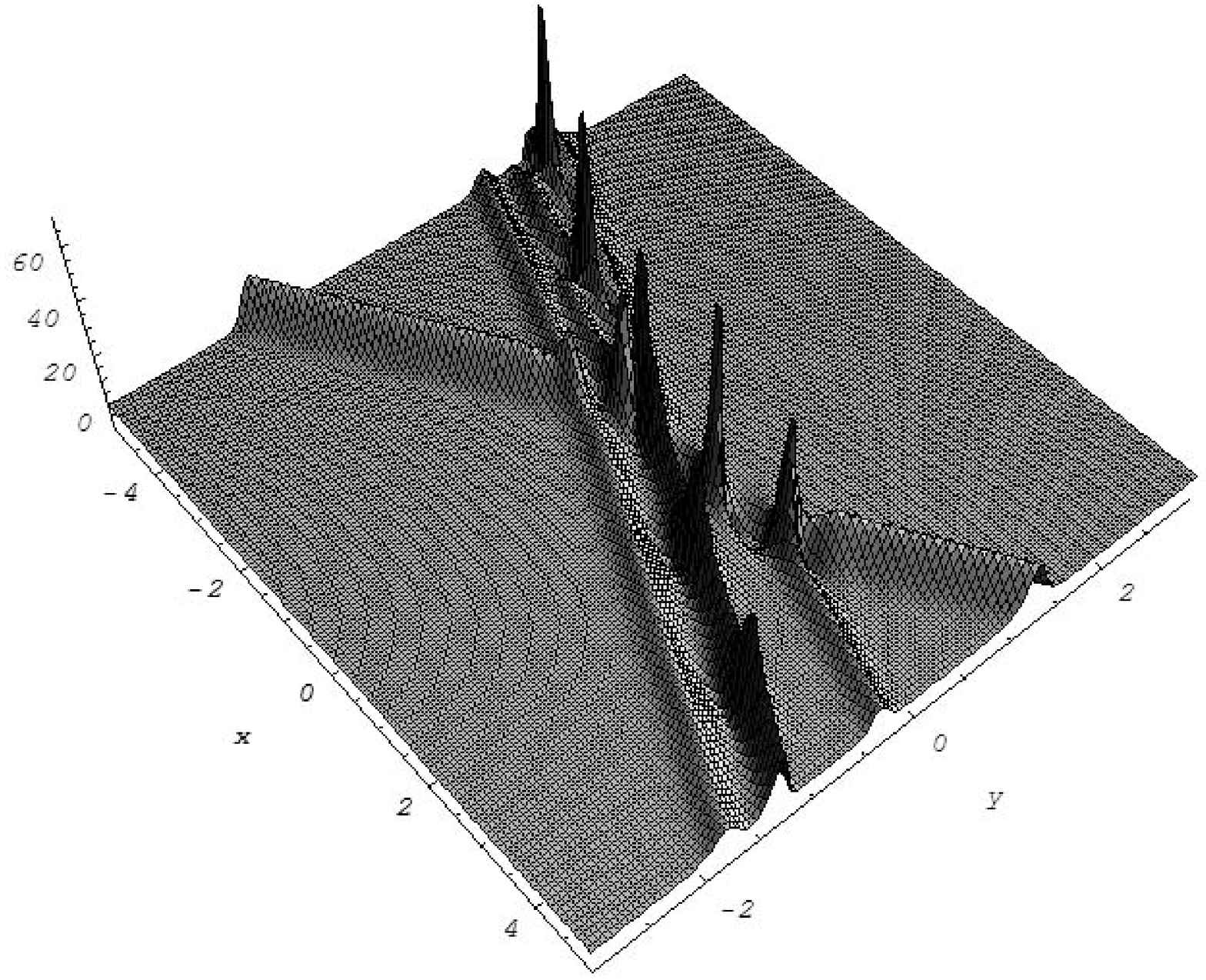}
\includegraphics[scale=0.3]{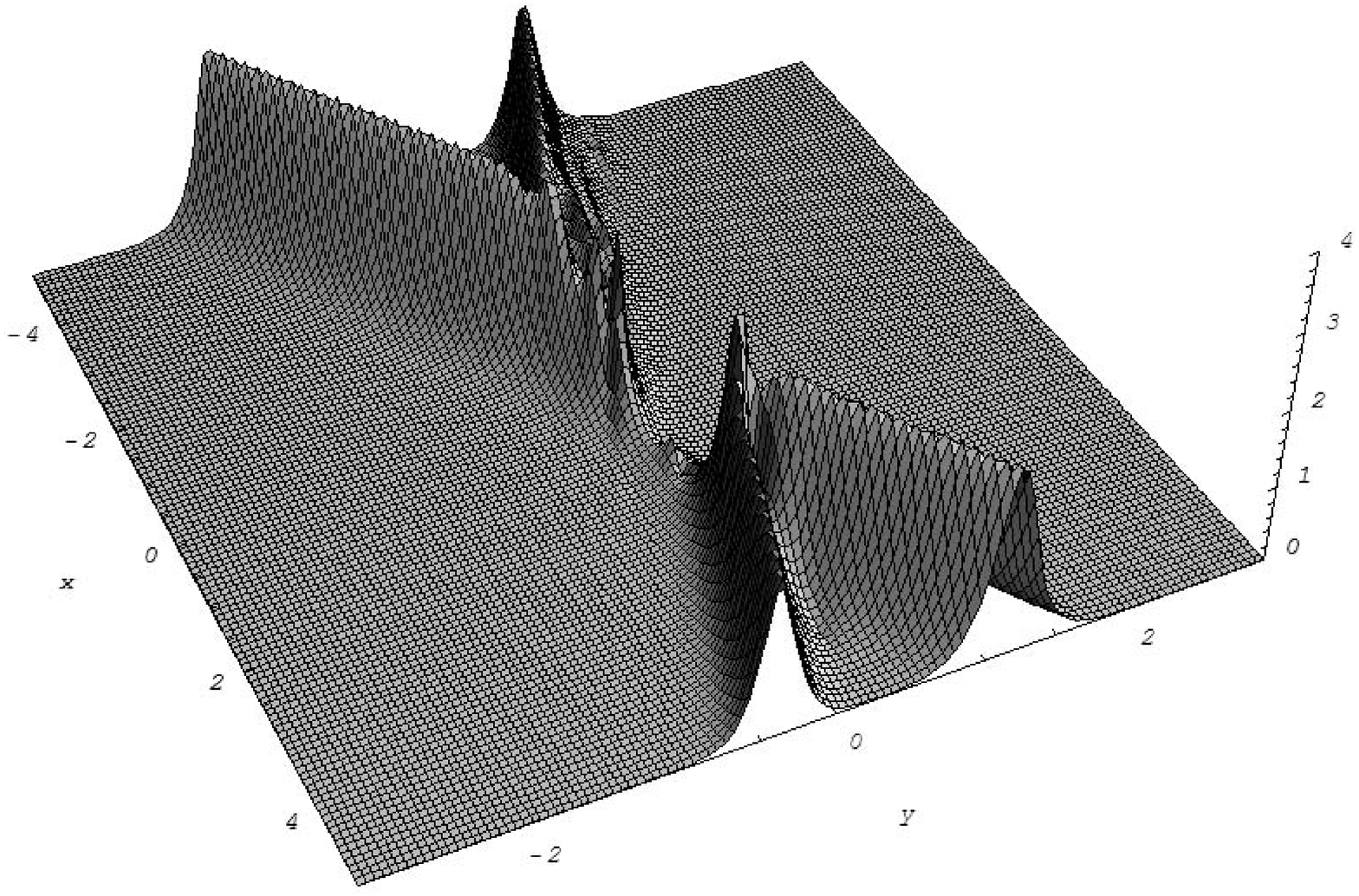}
}
\centerline{
\includegraphics[scale=0.32]{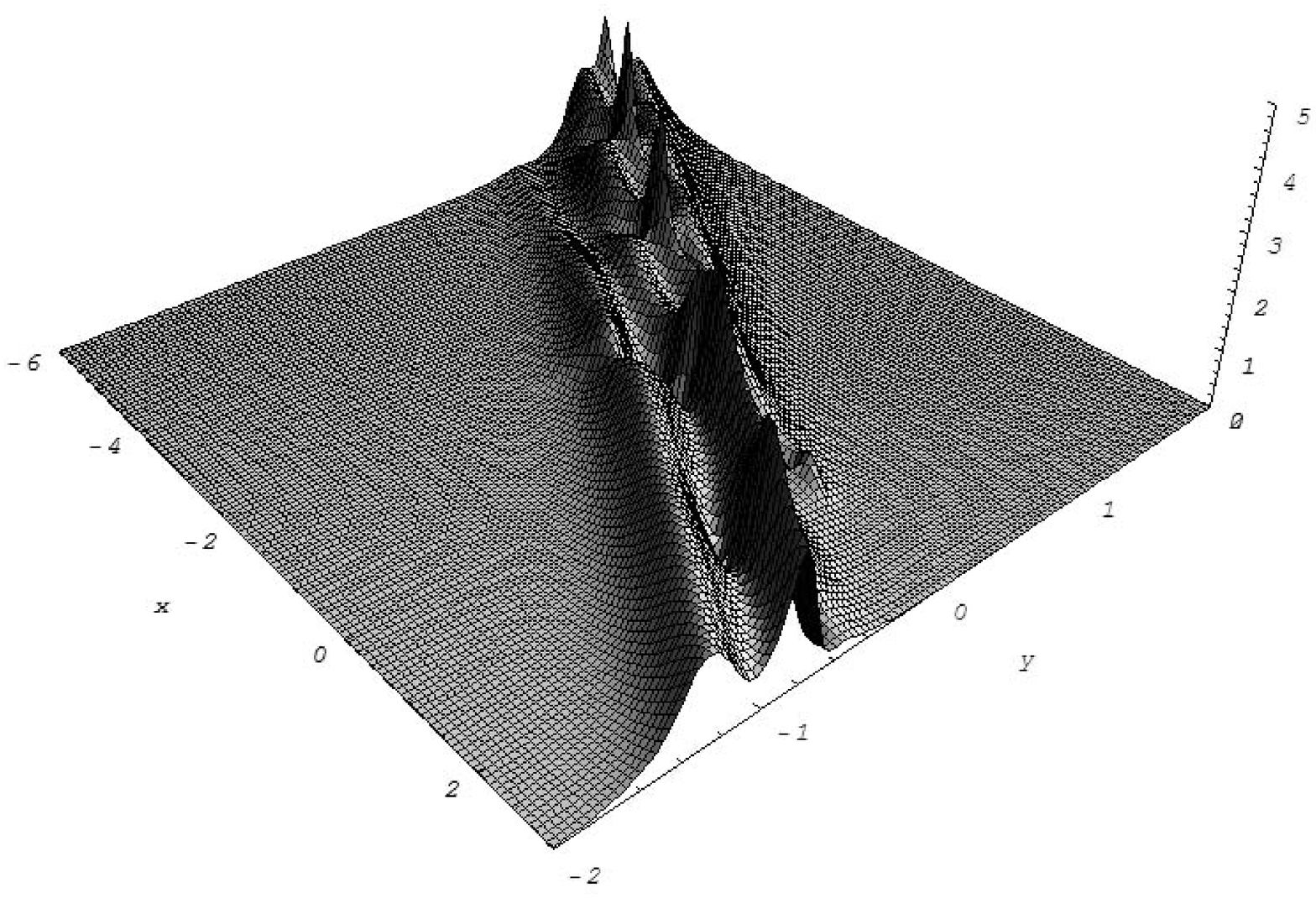}}
\caption{$(2,2,4)$-soliton solution. 
$p_1=2+3i$, $p_2=3-i$, 
$p_3=2-3i$, $p_4=3+i$, 
$s_1=2+2i$, $s_2=4+2i$,
$s_3=2-2i$, $s_4=-4-2i$, 
$q_1=2+i$, $q_2=2.01+i$, 
$q_3=2-i$, $q_4=2.01-i$, 
$r_1=1+i$, $r_2=1.5+i$, 
$r_3=1-i$, $r_4=1.5-i$, 
$\epsilon_1=\epsilon_2=\delta_1=\delta_2=-1$.
The top left graph is $-L$, the top right graph is
 $S^{(1)}$ and the bottom graph is $S^{(2)}$ at $t=0$.}
\label{fig3}
\end{figure}

Figures \ref{fig2} and \ref{fig3} show the $(2,2,4)$-soliton
interaction. In Fig.\ref{fig3}, an interesting soliton interaction 
of V-shape solitons and breather-type solitons is found. 
This interesting interaction is made by
the effect of the complicated condition for reality and complex
conjugacy. 
The similar interaction patterns were found in the case of one-component
\cite{oikawa}.

\section{Concluding Remarks}

We have derived two-component analogue of the
two-dimensional long wave-short wave interaction equations. 
Then, we have presented Wronskian solutions to the system. 
Interestingly, the direction of a soliton on $S^{(1)}$ is different from one
of a soliton on $S^{(2)}$. This is a special phenomenon in the case of
two-dimensional vector soliton. We also found the interesting soliton
interaction patterns including V-shape and breather-type soliton
solutions. 

Note that the generalization to $N$-component system is possible. 
In the $N$-component case, the solutions are constructed from
$(N+1)$-component Wronskian of the
$(N+1)$-component KP hierarchy. 

Recently, another two-dimensional analogue of long wave-short wave
interaction equations was derived asymptotically in \cite{kevre}. 
The integrability and the existence of $N$-soliton solution 
of this system is unknown. 
The study of multi-component generalization of this system  is also 
interesting. 

The detail of analysis of soliton interaction will 
be presented in the forthcoming
article.

\section*{Acknowledgements}

We are grateful to 
Dr. Takayuki Tsuchida for many helpful discussions.\break 
K.M.\ acknowledges support from Faculty Research Council at the
University of Texas-Pan American.

\section*{Appendix}

Consider the following three component Wronskian:
\begin{eqnarray*}
\fl &&\hskip-5mm\tau_{NML}=
\end{eqnarray*}
\tiny
\begin{eqnarray*}
\fl &&\left|\matrix{
\matrix{
\varphi_1 &\varphi_1^{(1)} &\cdots &\varphi_1^{(N-1)} \cr
\varphi_2 &\varphi_2^{(1)} &\cdots &\varphi_2^{(N-1)} \cr
\vdots &\vdots &&\vdots \cr
\varphi_{N+M+L} &\varphi_{N+M+L}^{(1)} &\cdots &\varphi_{N+M+L}^{(N-1)}}
\matrix{\vdots\cr\vdots\cr\vdots\cr\vdots}
\matrix{
\psi_1 &\psi_1^{(1)} &\cdots &\psi_1^{(M-1)} \cr
\psi_2 &\psi_2^{(1)} &\cdots &\psi_2^{(M-1)} \cr
\vdots &\vdots &&\vdots \cr
\psi_{N+M+L} &\psi_{N+M+L}^{(1)} &\cdots &\psi_{N+M+L}^{(M-1)}}
\matrix{\vdots\cr\vdots\cr\vdots\cr\vdots}
\matrix{
\chi_1 &\chi_1^{(1)} &\cdots &\chi_1^{(L-1)} \cr
\chi_2 &\chi_2^{(1)} &\cdots &\chi_2^{(L-1)} \cr
\vdots &\vdots &&\vdots \cr
\chi_{N+M+L} &\chi_{N+M+L}^{(1)} &\cdots &\chi_{N+M+L}^{(L-1)}}
}\right|
\end{eqnarray*}
\normalsize
where $\varphi_i$ are functions of $x_1$, $x_2$ and satisfy
\[
\partial_{x_2}\varphi_i=\partial_{x_1}^2\varphi_i\,,
\]
$\psi_i$ are arbitrary functions of $y_1$, 
$\chi_i$ are arbitrary functions of $z_1$, 
and 
$$
\varphi_i^{(n)}=\partial_{x_1}^n\varphi_i\,,
\qquad
\psi_i^{(n)}=\partial_{y_1}^n\psi_i\,,
\qquad
\chi_i^{(n)}=\partial_{z_1}^n\chi_i\,.
$$
This $\tau_{NML}$ satisfies
\begin{eqnarray*}
&&(D_{x_1}^2-D_{x_2})\tau_{N+1,M-1,L}\cdot\tau_{NML}=0\,,
\\
&&(D_{x_1}^2-D_{x_2})\tau_{N+1,M,L-1}\cdot\tau_{NML}=0\,,
\\
&&D_{x_1}D_{y_1}\tau_{NML}\cdot\tau_{NML}=2\tau_{N+1,M-1,L}\tau_{N-1,M+1,L}\,,
\\
&&D_{x_1}D_{z_1}\tau_{NML}\cdot\tau_{NML}=2\tau_{N+1,M,L-1}\tau_{N-1,M,L+1}\,.
\end{eqnarray*}
Setting
$$
f=\tau_{NML}\,,
\qquad
g=\tau_{N+1,M-1,L}\,,
\qquad
\bar g=\tau_{N-1,M+1,L}\,,
\qquad
h=\tau_{N+1,M,L-1}\,,
\qquad
\bar h=\tau_{N-1,M,L+1}\,,
$$
these $\tau$-functions satisfy
\begin{eqnarray*}
&&(D_{x_1}^2-D_{x_2})g\cdot f=0\,,
\\
&&(D_{x_1}^2+D_{x_2})\bar g\cdot f=0\,,
\\
&&(D_{x_1}^2-D_{x_2})h\cdot f=0\,,
\\
&&(D_{x_1}^2+D_{x_2})\bar h\cdot f=0\,,
\\
&&D_{x_1}(D_{y_1}+D_{z_1})f\cdot f=2(g\bar g+h\bar h)\,.
\end{eqnarray*}
Applying the transformation of the dependent variables
$$
S_1=\frac{g}{f}\,,
\qquad
\bar S_1=\frac{\bar g}{f}\,,
\qquad
S_2=\frac{h}{f}\,,
\qquad
\bar S_2=\frac{\bar h}{f}\,,
\qquad
L=-(2\log f)_{x_1x_1}\,,
$$
we obtain
\begin{eqnarray*}
&&\partial_{x_1}^2S_1-LS_1-\partial_{x_2}S_1=0\,,
\\
&&\partial_{x_1}^2\bar S_1-L\bar S_1+\partial_{x_2}\bar S_1=0\,,
\\
&&\partial_{x_1}^2S_2-LS_2-\partial_{x_2}S_2=0\,,
\\
&&\partial_{x_1}^2\bar S_2-L\bar S_2+\partial_{x_2}\bar S_2=0\,,
\\
&&-(\partial_{y_1}+\partial_{z_1})L=2(S_1\bar S_1+S_2\bar S_2)_{x_1}\,,
\end{eqnarray*}
Applying the change of independent variables
$$
x_1=x\,,
\qquad
x_2=-iy\,,
\qquad
y_1=y-t\,,
\qquad
z_1=y-t\,,
$$
i.e., 
$$
\partial_x=\partial_{x_1}\,,
\qquad
\partial_y=\partial_{y_1}+\partial_{z_1}-i\partial_{x_2}\,,
\qquad
\partial_t=-\partial_{y_1}-\partial_{z_1}\,,
$$
we obtain
\begin{eqnarray*}
&&\partial_x^2S_1-LS_1-i(\partial_t+\partial_y)S_1=0\,,
\\
&&\partial_x^2\bar S_1-L\bar S_1+i(\partial_t+\partial_y)\bar S_1=0\,,
\\
&&\partial_x^2S_2-LS_2-i(\partial_t+\partial_y)S_2=0\,,
\\
&&\partial_x^2\bar S_2-L\bar S_2+i(\partial_t+\partial_y)\bar S_2=0\,,
\\
&&L_t=2(S_1\bar S_1+S_2\bar S_2)_x\,.
\end{eqnarray*}

In the above solution, we consider the replacements of $N, M$ and $L$ by  
\[
N\to 2N\,,
\qquad
M\to N\,,
\qquad
L\to N\,,
\]
i.e., 
consider
$$
f=\tau_{2N,N,N}
\qquad
g=\tau_{2N+1,N-1,N}
\qquad
\bar g=\tau_{2N-1,N+1,N}
\qquad
h=\tau_{2N+1,N,N-1}
\qquad
\bar h=\tau_{2N-1,N,N+1}\,.
$$
Setting
$$
\psi_{2N+1}=\psi_{2N+2}=\cdots=\psi_{4N}=0\,,\quad 
\chi_1=\chi_2=\cdots=\chi_{2N}=0\,,
$$
we have
\small
\begin{eqnarray*}
\fl &&\hskip-5mm\tau_{2N+n+m,N-n,N-m}=
\\
\fl &&\left|\matrix{
\matrix{
\varphi_1 &\varphi_1^{(1)} &\cdots &\varphi_1^{(2N+n+m-1)} \cr
\vdots &\vdots &&\vdots \cr
\varphi_{2N} &\varphi_{2N}^{(1)} &\cdots &\varphi_{2N}^{(2N+n+m-1)} \cr
\noalign{\vskip4pt}
\varphi_{2N+1} &\varphi_{2N+1}^{(1)} &\cdots &\varphi_{2N+1}^{(2N+n+m-1)} \cr
\vdots &\vdots &&\vdots \cr
\varphi_{4N} &\varphi_{4N}^{(1)} &\cdots &\varphi_{4N}^{(2N+n+m-1)}}
\matrix{\vdots\cr\vdots\cr\vdots\cr\vdots\cr\vdots\cr\vdots}
\matrix{
\psi_1 &\psi_1^{(1)} &\cdots &\psi_1^{(N-n-1)} \cr
\vdots &\vdots &&\vdots \cr
\psi_{2N} &\psi_{2N}^{(1)} &\cdots &\psi_{2N}^{(N-n-1)} \cr
\noalign{\vskip1pt}\multispan{4}{\dotfill} \cr
&&&\vphantom{\psi_{2N+1}^{(N-n-1)}} \cr
&&{\bf 0}&\vphantom{\vdots} \cr
&&&\vphantom{\psi_{4N}^{(N-n-1)}}}
\matrix{\vdots\cr\vdots\cr\vdots\cr\vdots\cr\vdots\cr\vdots}
\matrix{
&&&\vphantom{\chi_1^{(N-m-1)}} \cr
&&&\vphantom{\chi_1^{(N-m-1)}} \cr
&&{\bf 0}& \cr
&&&\vphantom{\chi_{2N}^{(N-m-1)}} \cr
\noalign{\vskip1pt}\multispan{4}{\dotfill} \cr
\chi_{2N+1} &\chi_{2N+1}^{(1)} &\cdots &\chi_{2N+1}^{(N-m-1)} \cr
\vdots &\vdots &&\vdots \cr
\chi_{4N} &\chi_{4N}^{(1)} &\cdots &\chi_{4N}^{(N-m-1)}}
}\right|\,.
\end{eqnarray*}
\normalsize
Note that this determinant is 0 if $N-n>2N$ or $N-m>2N$. 
Consider the following table of non-zero $\tau$-functions:
$$
\matrix{
&m=N &&
 &\tau_{2N,2N,0} &\tau_{2N+1,2N-1,0} &\cdots
 &\tau_{4N-1,1,0} &\tau_{4N,0,0} \cr
&m=N-1 &&
 &\tau_{2N-1,2N,1} &\tau_{2N,2N-1,1} &\cdots
 &\tau_{4N-2,1,1} &\tau_{4N-1,0,1} \cr
&&&
 &\vdots &\vdots &&\vdots &\vdots \cr
&m=-N+1 &&
 &\tau_{1,2N,2N-1} &\tau_{2,2N-1,2N-1} &\cdots
 &\tau_{2N,1,2N-1} &\tau_{2N+1,0,2N-1} \cr
&m=-N &&
 &\tau_{0,2N,2N} &\tau_{1,2N-1,2N} &\cdots
 &\tau_{2N-1,1,2N} &\tau_{2N,0,2N} \cr
\cr
&&&
 &n=-N &n=-N+1 &&n=N-1 &n=N}
$$
The $\tau$-function in the center on the table is 
corresponding to $f$:
$$
\matrix{
\tau_{2N,2N,0} &\tau_{2N+1,2N-1,0} &\cdots &\cdots &\cdots
 &\tau_{4N-1,1,0} &\tau_{4N,0,0} \cr
\tau_{2N-1,2N,1} &\tau_{2N,2N-1,1} &\cdots &\cdots &\cdots
 &\tau_{4N-2,1,1} &\tau_{4N-1,0,1} \cr
\vdots &\vdots & &h & &\vdots &\vdots \cr
\vdots &\vdots &\bar g &f &g &\vdots &\vdots \cr
\vdots &\vdots & &\bar h & &\vdots &\vdots \cr
\tau_{1,2N,2N-1} &\tau_{2,2N-1,2N-1} &\cdots &\cdots &\cdots
 &\tau_{2N,1,2N-1} &\tau_{2N+1,0,2N-1} \cr
\tau_{0,2N,2N} &\tau_{1,2N-1,2N} &\cdots &\cdots &\cdots
 &\tau_{2N-1,1,2N} &\tau_{2N,0,2N}}
$$
Now we want to find the condition of complex conjugacy 
($f$: real, $\bar g=g^*$, $\bar h=h^*$ where $^*$ means complex
conjugate). 
For the function $\tau_{2N+n+m,N-n,N-m}$, 
the bilinear equations of two-dimensional 
Toda lattice
\begin{eqnarray*}
&&D_{x_1}D_{y_1}\tau_{2N+n+m,N-n,N-m}\cdot\tau_{2N+n+m,N-n,N-m}\\
&& \qquad \quad =2\tau_{2N+n+m+1,N-n-1,N-m}\tau_{2N+n+m-1,N-n+1,N-m}
\\
 &&D_{x_1}D_{z_1}\tau_{2N+n+m,N-n,N-m}\cdot\tau_{2N+n+m,N-n,N-m}\\
&& \qquad \quad =2\tau_{2N+n+m+1,N-n,N-m-1}\tau_{2N+n+m-1,N-n,N-m+1}
\end{eqnarray*}
are satisfied. 
Since the function $\tau_{4N,0,0}$ depends on only 
$x_1$, $x_2$, the following bilinear equations are satisfied: 
\begin{eqnarray*}
&&D_{x_1}D_{y_1}\frac{\tau_{2N+n+m,N-n,N-m}}{\tau_{4N,0,0}}
\cdot\frac{\tau_{2N+n+m,N-n,N-m}}{\tau_{4N,0,0}}\\
&&\quad =2\frac{\tau_{2N+n+m+1,N-n-1,N-m}}{\tau_{4N,0,0}}
\frac{\tau_{2N+n+m-1,N-n+1,N-m}}{\tau_{4N,0,0}}\,,
\\
&&D_{x_1}D_{z_1}\frac{\tau_{2N+n+m,N-n,N-m}}{\tau_{4N,0,0}}
\cdot\frac{\tau_{2N+n+m,N-n,N-m}}{\tau_{4N,0,0}}\\
&&\quad =2\frac{\tau_{2N+n+m+1,N-n,N-m-1}}{\tau_{4N,0,0}}
\frac{\tau_{2N+n+m-1,N-n,N-m+1}}{\tau_{4N,0,0}}\,.
\end{eqnarray*}
The table of solutions of these bilinear equations are as follows:
$$
\matrix{
\displaystyle\frac{\tau_{2N,2N,0}}{\tau_{4N,0,0}}
 &\displaystyle\frac{\tau_{2N+1,2N-1,0}}{\tau_{4N,0,0}}
 &\cdots &\cdots &\cdots
 &\displaystyle\frac{\tau_{4N-1,1,0}}{\tau_{4N,0,0}}
 &1 \cr
\displaystyle\frac{\tau_{2N-1,2N,1}}{\tau_{4N,0,0}}
 &\displaystyle\frac{\tau_{2N,2N-1,1}}{\tau_{4N,0,0}}
 &\cdots &\cdots &\cdots
 &\displaystyle\frac{\tau_{4N-2,1,1}}{\tau_{4N,0,0}}
 &\displaystyle\frac{\tau_{4N-1,0,1}}{\tau_{4N,0,0}} \cr
\vdots &\vdots & &\displaystyle\frac{h}{\tau_{4N,0,0}} & &\vdots &\vdots \cr
\vdots &\vdots &\displaystyle\frac{\bar g}{\tau_{4N,0,0}}
 &\displaystyle\frac{f}{\tau_{4N,0,0}}
 &\displaystyle\frac{g}{\tau_{4N,0,0}} &\vdots &\vdots \cr
\vdots &\vdots & &\displaystyle\frac{\bar h}{\tau_{4N,0,0}}
 & &\vdots &\vdots \cr
\displaystyle\frac{\tau_{1,2N,2N-1}}{\tau_{4N,0,0}}
 &\displaystyle\frac{\tau_{2,2N-1,2N-1}}{\tau_{4N,0,0}}
 &\cdots &\cdots &\cdots
 &\displaystyle\frac{\tau_{2N,1,2N-1}}{\tau_{4N,0,0}}
 &\displaystyle\frac{\tau_{2N+1,0,2N-1}}{\tau_{4N,0,0}} \cr
\displaystyle\frac{\tau_{0,2N,2N}}{\tau_{4N,0,0}}
 &\displaystyle\frac{\tau_{1,2N-1,2N}}{\tau_{4N,0,0}}
 &\cdots &\cdots &\cdots
 &\displaystyle\frac{\tau_{2N-1,1,2N}}{\tau_{4N,0,0}}
 &\displaystyle\frac{\tau_{2N,0,2N}}{\tau_{4N,0,0}}}
$$
Since the function $\tau_{0,2N,2N}$ does not depend on $x_1$, 
the following bilinear equations are also satisfied: 
\begin{eqnarray*}
&&D_{x_1}D_{y_1}\frac{\tau_{2N+n+m,N-n,N-m}}{\tau_{0,2N,2N}}
\cdot\frac{\tau_{2N+n+m,N-n,N-m}}{\tau_{0,2N,2N}}\\
&&\quad =2\frac{\tau_{2N+n+m+1,N-n-1,N-m}}{\tau_{0,2N,2N}}
\frac{\tau_{2N+n+m-1,N-n+1,N-m}}{\tau_{0,2N,2N}}\,,
\\
&&D_{x_1}D_{z_1}\frac{\tau_{2N+n+m,N-n,N-m}}{\tau_{0,2N,2N}}
\cdot\frac{\tau_{2N+n+m,N-n,N-m}}{\tau_{0,2N,2N}}\\
&&\quad =2\frac{\tau_{2N+n+m+1,N-n,N-m-1}}{\tau_{0,2N,2N}}
\frac{\tau_{2N+n+m-1,N-n,N-m+1}}{\tau_{0,2N,2N}}\,.
\end{eqnarray*}
The table of solutions of these bilinear equations are as follows:
$$
\matrix{
\displaystyle\frac{\tau_{2N,2N,0}}{\tau_{0,2N,2N}}
 &\displaystyle\frac{\tau_{2N+1,2N-1,0}}{\tau_{0,2N,2N}}
 &\cdots &\cdots &\cdots
 &\displaystyle\frac{\tau_{4N-1,1,0}}{\tau_{0,2N,2N}}
 &\displaystyle\frac{\tau_{4N,0,0}}{\tau_{0,2N,2N}} \cr
\displaystyle\frac{\tau_{2N-1,2N,1}}{\tau_{0,2N,2N}}
 &\displaystyle\frac{\tau_{2N,2N-1,1}}{\tau_{0,2N,2N}}
 &\cdots &\cdots &\cdots
 &\displaystyle\frac{\tau_{4N-2,1,1}}{\tau_{0,2N,2N}}
 &\displaystyle\frac{\tau_{4N-1,0,1}}{\tau_{0,2N,2N}} \cr
\vdots &\vdots & &\displaystyle\frac{h}{\tau_{0,2N,2N}} & &\vdots &\vdots \cr
\vdots &\vdots &\displaystyle\frac{\bar g}{\tau_{0,2N,2N}}
 &\displaystyle\frac{f}{\tau_{0,2N,2N}}
 &\displaystyle\frac{g}{\tau_{0,2N,2N}} &\vdots &\vdots \cr
\vdots &\vdots & &\displaystyle\frac{\bar h}{\tau_{0,2N,2N}}
 & &\vdots &\vdots \cr
\displaystyle\frac{\tau_{1,2N,2N-1}}{\tau_{0,2N,2N}}
 &\displaystyle\frac{\tau_{2,2N-1,2N-1}}{\tau_{0,2N,2N}}
 &\cdots &\cdots &\cdots
 &\displaystyle\frac{\tau_{2N,1,2N-1}}{\tau_{0,2N,2N}}
 &\displaystyle\frac{\tau_{2N+1,0,2N-1}}{\tau_{0,2N,2N}} \cr
1
 &\displaystyle\frac{\tau_{1,2N-1,2N}}{\tau_{0,2N,2N}}
 &\cdots &\cdots &\cdots
 &\displaystyle\frac{\tau_{2N-1,1,2N}}{\tau_{0,2N,2N}}
 &\displaystyle\frac{\tau_{2N,0,2N}}{\tau_{0,2N,2N}}}
$$
Consider the case in which the relations
$$
\frac{\tau_{4N-1,1,0}}{\tau_{4N,0,0}}
=\left(\frac{\tau_{1,2N-1,2N}}{\tau_{0,2N,2N}}\right)^*\,,
$$
and 
$$
\frac{\tau_{4N-1,0,1}}{\tau_{4N,0,0}}
=\left(\frac{\tau_{1,2N,2N-1}}{\tau_{0,2N,2N}}\right)^*\,,
$$
are satisfied. 
If we choose elements in 
the solution in the section
3 as $\varphi_i$, $\psi_i$, $\chi_i$, these relations are satisfied, and 
we also notice that the relations
$$
\frac{f}{\tau_{4N,0,0}}
=\left(\frac{f}{\tau_{0,2N,2N}}\right)^*\,,
$$
$$
\frac{g}{\tau_{4N,0,0}}
=\left(\frac{\bar g}{\tau_{0,2N,2N}}\right)^*\,,
$$
$$
\frac{h}{\tau_{4N,0,0}}
=\left(\frac{\bar h}{\tau_{0,2N,2N}}\right)^*\,,
$$
are satisfied. 
From these relations, we conclude that 
the complex conjugacy condition 
$$
\left(\frac{g}{f}\right)^*=\frac{\bar g}{f}\,,
\qquad
\left(\frac{h}{f}\right)^*=\frac{\bar h}{f}
\,,$$
$$
f \mathcal{G}  \hbox{ : real}\,,
$$
where $\mathcal{G}$ is an exponential factor which is a gauge function,
is satisfied.


\end{document}